\documentclass[twocolumn]{raa_twocolumn}

\usepackage{graphicx,times}
\usepackage{natbib}
\usepackage{amssymb,amsmath}
\usepackage{enumitem}
\bibpunct{(}{)}{;}{a}{}{,}
\usepackage[pagebackref=true]{hyperref}

\begin{document}
   \title{Mock Observations for the CSST Mission: Main Surveys -- the Mock Catalogue}
   \volnopage{ {\bf 20XX} Vol.\ {\bf X} No. {\bf XX}, 000--000}
   \setcounter{page}{1}

    \author{Cheng-Liang Wei\inst{1}, Yu Luo\inst{2}, Hao Tian\inst{3}, Ming Li\inst{3}, Yi-Sheng Qiu\inst{4,5}, Guo-Liang Li\inst{1}, Yue-Dong Fang\inst{6}, Xin Zhang\inst{3}, De-Zi Liu\inst{7}, Nan Li\inst{3}, Ran Li\inst{3,8}, Huan-Yuan Shan\inst{9}, Lin Nie\inst{9}, Zizhao He\inst{1}, Lei Wang\inst{1}, Xi Kang\inst{4}, Dongwei Fan\inst{3}, Yang Chen\inst{10}, Xiaoting Fu\inst{1}, Chao Liu\inst{3} 
    }

    \institute{Purple Mountain Observatory, Chinese Academy of Sciences, 10 Yuanhua Road, Nanjing 210023, People’s Republic of China;
    \and School of Physics and Electronics, Hunan Normal University, 36 Lushan Road, Changsha 410081,People’s Republic of China; {\it luoyupmo@gmail.com}
 	\and Key Laboratory of Space Astronomy and Technology, National Astronomical Observatories, Chinese Academy of Sciences, 20A Datun Road, Beijing 100101, People’s Republic of China; {\it tianhao@nao.cas.cn}
 	\and Zhejiang University–Purple Mountain Observatory Joint Research Center for Astronomy, Zhejiang University, Hangzhou 310027, People's Republic of China
 	\and Research Center for Astronomical Computing, Zhejiang Laboratory, Hangzhou 311121, People’s Republic of China
	\and Universit\"ats-Sternwarte M\"unchen, Fakult\"at f\"ur Physik, Ludwig-Maximilians-Universit\"at M\"unchen, Scheinerstrasse 1, 81679 M\"unchen, Germany
	\and South-Western Institute for Astronomy Research, Yunnan University, Kunming, Yunnan, 650500, People’s Republic of China
	\and School of Physics and Astronomy, Beijing Normal University, Beijing 100875, People’s Republic of China
	\and Shanghai Astronomical Observatory, Chinese Academy of Sciences, Shanghai 200030, People’s Republic of China
	\and School of Physics and Optoelectronic Engineering, Anhui University, Hefei, 230601, People’s Republic of China
	}

\abstract{The Chinese Space Station Survey Telescope (CSST) is a flagship space mission, supported by the China Manned Space Project, designed to carry out a large-area sky survey to explore the nature of dark matter and dark energy in the Universe. The onboard multi-band imaging and slitless spectroscopic modules will enable us to obtain photometric data for billions of galaxies and stars, as well as hundreds of millions of spectroscopic measurements, advancing various scientific analyses such as galaxy clustering and weak gravitational lensing. To support the image simulations for the main survey of the CSST mission, we present a mock catalogue of stars and galaxies. For stars, the mock catalogue is generated using either {\tt Galaxia} or {\tt TRILEGAL}, both of which provide a range of stellar properties to meet the requirements of CSST image simulations. For galaxies, we built a mock light-cone up to redshift $z \sim 3.5$ from the cosmological $N$-body simulation and populated the mock galaxy catalogue from the dark mater haloes using a semi-analytical
galaxy formation model. We then performed a full-sky ray-tracing simulation of weak gravitational lensing to obtain lensing shear at the position of each galaxy in the light-cone. To support both multi-band imaging and slitless spectroscopic simulations, we computed the spectral energy distribution (SED) for each galaxy based on its star formation history using a supervised deep-learning model and determined the magnitudes in each band using the CSST throughputs. Finally, the properties of our mock galaxies include positions, redshifts, stellar masses, shapes, sizes, SEDs, lensing shears and magnifications. We have validated our mock catalogue against observational data and theoretical models, with results showing good overall agreement. The catalogue provides a flexible dataset for the development of CSST image processing and can support a wide range of cosmological analyses within the CSST mission.
}

   \authorrunning{Wei et al.}
   \titlerunning{Mock Catalogue}
   \maketitle
%________________________________________________ sections below
\section{Introduction}           %% first-level sections will be auto-capitalized
\label{sect:intro}
To accurately understand the evolution of matter in the Universe, astronomers rely to large-area galaxy surveys to study the growth of cosmic structures and uncover the nature of the accelerated expansion of the universe (\citealt{Fu+2014MNRAS_cfhtlens, Asgari+2021A&A_Kids1000, Doux+2022MNRAS_DES, Dalal+2023PhRvD_HSC}). Current and upcoming galaxy surveys, such as the Vera C. Rubin Observatory (LSST, \citealt{Abell+2009arXiv_LSST, Ivezi+2019ApJ_LSST}) and Euclid (\citealt{Laureijs+2011arXiv_Euclid, Mellier+2024arXiv_Euclid}), will generate vast amounts of high-quality data, enabling unprecedented precision in constraining cosmological parameters, especially the equation of state of dark energy (\citealt{Mandelbaum+2018ARA&A, Zhang+2022MNRAS, Fanizza+2023mgm}). 

The Chinese Space Station Survey Telescope (CSST), a flagship project of the Stage-IV galaxy surveys, will carry out a 10-year space mission to perform a large-area galaxy survey with $\sim$17,500 $\deg^2$ of wide field and $\sim$400 $\deg^2$ of deep field \citep{Zhan+2021CSB}. CSST will determine galaxy shapes by imaging billions of galaxies within 7-bands ($NUV$, $u$, $g$, $r$, $i$, $z$, $y$), reaching a magnitude limit of $m_g \sim 26.3$ in wide field and $\sim 1.2$ magnitudes fainter in the deep field. Simultaneously, its slitless spectroscopic module will provide low-resolution spectra for hundreds of millions of galaxies to determine their redshifts (as outlined in the CSST science book). These rich datasets will support precise analyses of galaxy clustering and weak gravitational lensing, delivering robust cosmological constraints (\citealt{Liu+2023A&A, Miao+2023MNRAS, Song+2024ApJ, Xiong+2024arXiv}).

As statistical accuracy improves in these Stage-IV surveys, systematic errors will be critically important for maximizing the scientific returns of these current and future galaxy surveys (\citealt{Salvati+2020A&A, Yao+2024MNRAS}). It is essential to develop an end-to-end imaging simulator to optimize the data processing and conduct preliminary scientific analyses (\citealt{Plazas+2020Symm, Sanchez+2020MNRAS_LSST, Serrano+2024A&A_Euclid}). For CSST, so far, a public release of the imaging simulator\footnote{\url{https://csst-tb.bao.ac.cn/code/csst-sims/csst_msc_sim}} (version 3.1.0) has been developed by the CSST scientific data processing and analysis system to simulate multi-band imaging and slitless spectroscopic observations (Wei \& Fang et al. 2025, submitted). To produce highly realistic mock images, an optical emulator has been developed to simulate high-fidelity PSFs of CSST (Ban et al. 2025, submitted) and various sources of noise have been included, such as shot noise, sky background, and comprehensive detector effects. To achieve this, we utilize {\tt Galsim}\footnote{\url{https://github.com/GalSim-developers/GalSim}} (\citealt{Rowe+2015A&C_galsim}) to generate photons from a given object, taking into account the throughputs of the CSST filter system. These throughputs encompass the mirror efficiency, filter transmission, and quantum efficiency of the detector. For more details of CSST imaging simulator we refer the reader to Wei \& Fang et al. (2025, submitted). With the CSST imaging simulator, we have released a set of $50 \deg^2$ data products, which has been used to develop the pipeline of data processing and optimize the survey design of CSST.

This paper provids a detailed description of the input mock catalogue used in the CSST main survey simulation. The structure is organized as follows. An overview of the input mock catalogue is shown in Sec. 2. In Sec. 3 \& Sec. 4, we describe the production and primary characteristics of the star and galaxy catalogues, respectively. Sec. 5 presents the computation of weak lensing properties by a full-sky ray-tracing simulation.  Sec. 6 details the validation of the properties of the mock catalogue against observational constraints and theoretical models. Although tailored for the CSST simulation, the catalogue can be very useful for various studies due to its comprehensive inclusion of galaxy properties and consistently calculated weak lensing observables down to sub-arcminute scales. Finally, Sec. 7 provides a summary of our mock catalogue.

\begin{figure*}
    \centering
    \includegraphics[width=0.9\textwidth]{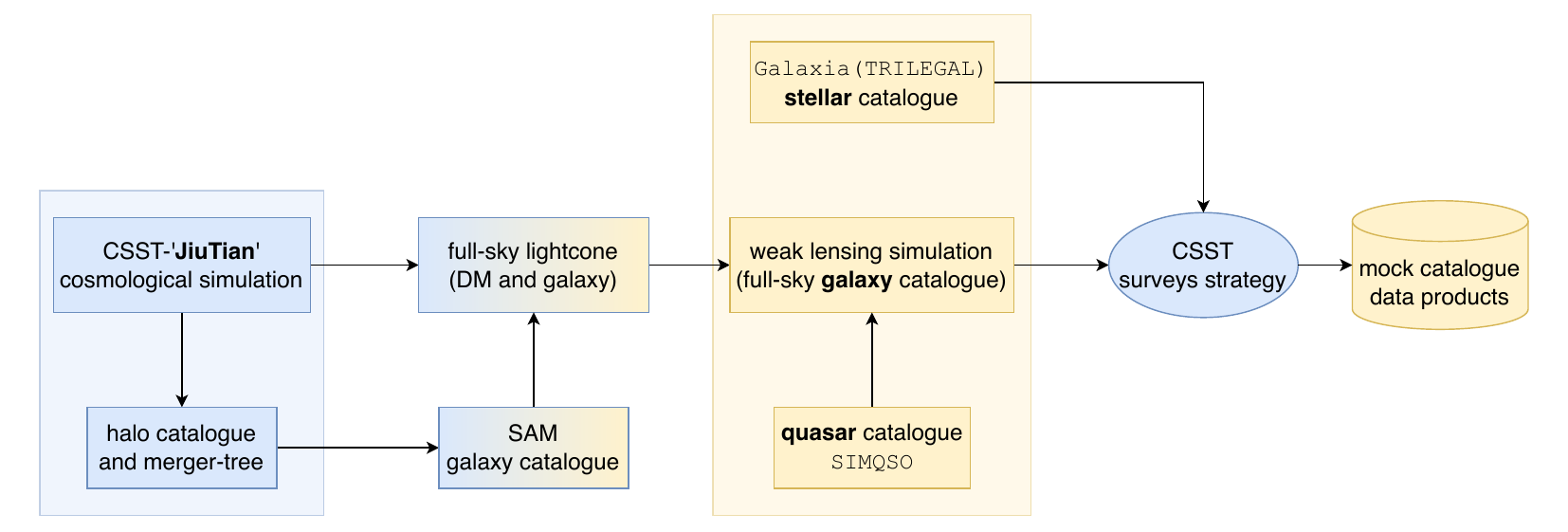}
    \caption{An overview of the pipline for the input mock catalogue.}
    \label{fig:catsFlow}
\end{figure*}

\section{Overview of the mock catalogue}
The input mock catalogue for the recent release comprises stars, galaxies and quasars (Fig.~\ref{fig:catsFlow}).  To support the development of imaging simulator and various scientific studies, it is of great importance that the catalogue should include a range of realistic features. For stars, we generate the field star catalogue on demand using either {\tt Galaxia} \citep{Sharma2011ApJ...730....3S} or {\tt TRILEGAL} \citep{Girardi+2012ASSP_TRILEGAL}, two widely used stellar population synthesis models for the Milky Way. Both models can generate mock star catalogues that effectively support the general requirements of CSST image simulation. For galaxies, several methods can be employed to generate high-precision galaxy catalogues (\citealt{Merson+2013MNRAS_SAM,     Carretero+2015MNRAS_MICE, Torrey+2015MNRAS_Illustris, Barrera+2023MNRAS_MillenniumTNG}). Hydrodynamical simulations, which model complex baryonic feedback processes within galaxies, can directly produce galaxy samples with detailed physical properties, such as morphology and star formation rates (\citealt{Vogelsberger+2014MNRAS_Illustris, Kaviraj+2017MNRAS_HorizonAGN, Schaye+2023MNRAS_Flamingo}). However, current hydrodynamical simulations are computationally expensive and rely on subgrid physics models (\citealt{Crain+2015MNRAS_Eagle, Habouzit+2021MNRAS}). A more efficient and widely used alternative is based on high-precision N-body simulation. This approach, adopted in cosmoDC2 (\citealt{Korytov+2019ApJS_cosmoDC2}) and Flagship (\citealt{Castander+2024arXiv_Flagship}), involves constructing dark matter light-cones using large-volume and high-mass resolution cosmological simulations. Dark matter haloes are identified within the light-cone and galaxies are populated using algorithms such as semi-analytical galaxy formation models (SAM) or halo occupation distribution models (HOD).

Regarding CSST, we have conducted the `JiuTian' simulations \citep{Jiutian_2025arXiv250321368H}, a series of cosmological N-body simulations with varying box sizes and resolutions, designed to support mission optimization and scientific analysis preparation. Given the wide-field and deep observation capabilities of CSST, the mass of dark matter particle in the simulation should be $\sim 10^8 {\rm M}_{\odot}$ to support studies of large-scale structure and weak gravitational lensing. In particular, the JiuTian-1G simulation (hereafter JT1G), with a box size of $1 h^{-1}{\rm Gpc}$, is utilized to construct a full-sky light-cone covering redshifts up to $z_{\rm s} \sim 3.5$ for CSST imaging simulator. Galaxies are generated using SAM (\citealt{Luo+2016MNRAS}) and weak lensing signals are incorporated through a full-sky ray-tacing simulation(\citealt{Wei+2018ApJ}). Specifically, to meet the needs of spectral analysis, {\tt STARDUSTER}\footnote{\url{https://github.com/yqiuu/starduster}} (\citealt{Qiu+2022ApJ}) has been used to generate the spectral energy distribution (SED) for each galaxy in the catalogue. Below, we provide a detailed description of the input mock catalogues used in CSST imaging simulator.

\begin{figure}
    \centering
    \includegraphics[width=0.45\textwidth]{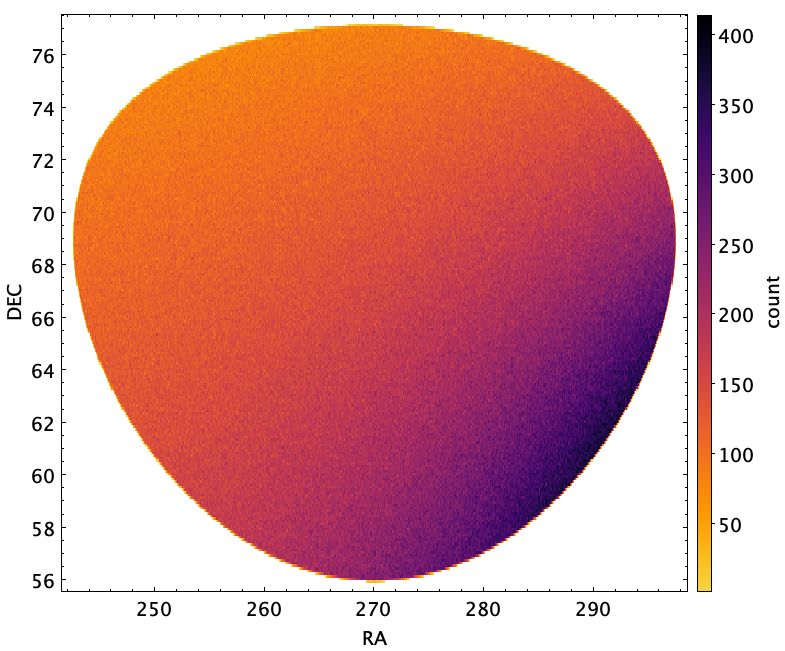}
    \caption{The density distribution of the field stars around the direction of $(\alpha, \delta)=(270^\circ, 67^\circ)$.}
    \label{fig:star_ra_dec}
\end{figure}

\begin{figure*}
    \centering
    \includegraphics[width=\textwidth]{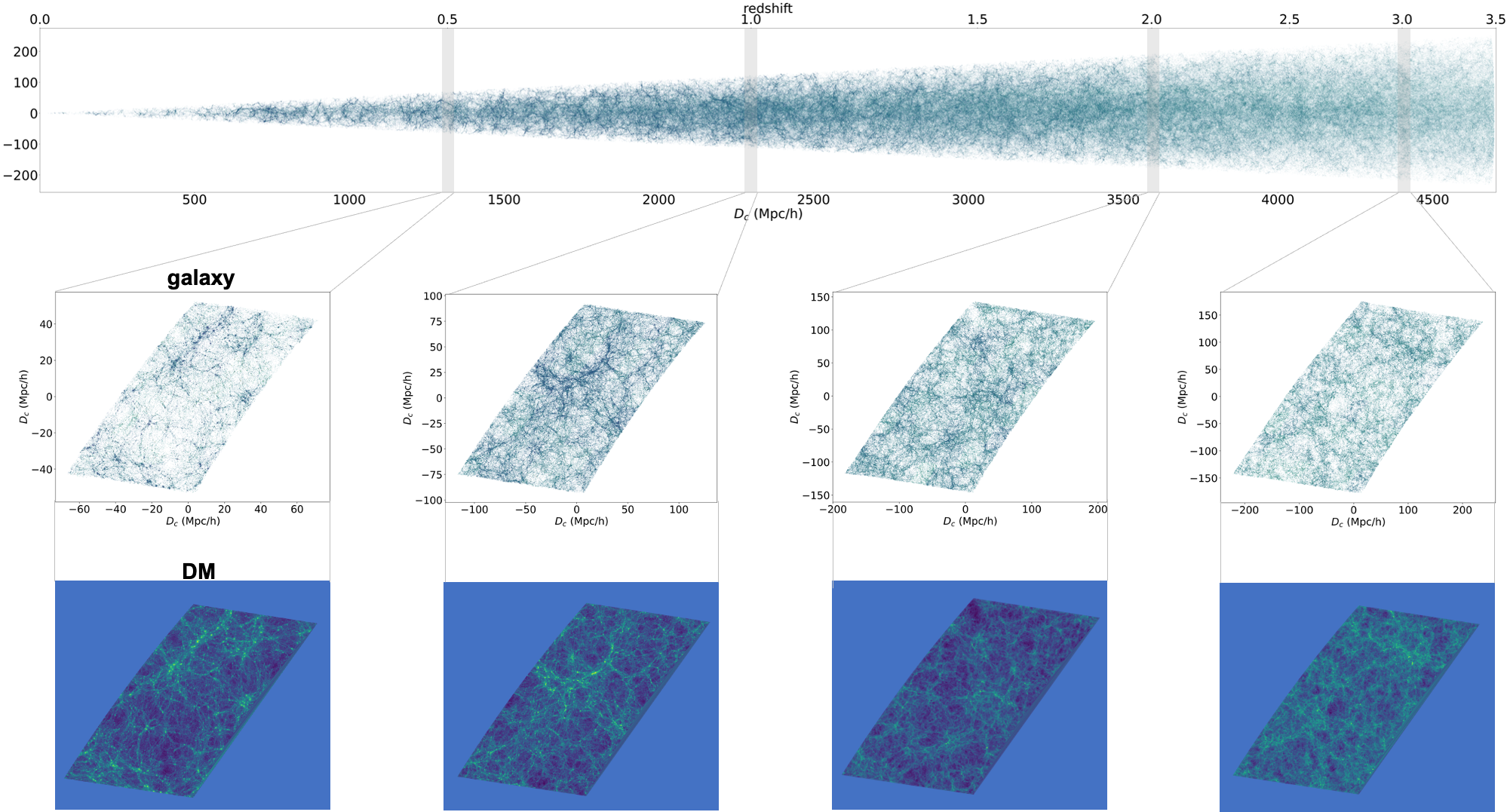}
    \caption{In the top panel, we show a sky region of $\sim 13$ ${\rm deg}^2$ from the full-sky galaxy light-cone. The middle and bottom panels illustrate the distribution of mock galaxies and dark matter density fields, respectively, at redshifts of 0.5, 1.0, 2.0 and 3.0, where the filaments of large-scale structure can be clearly distinguished in detail.}
    \label{fig:lightcone}
\end{figure*}

\section{star Catalogue}
Field stars from the Milky Way are essential for the imaging simulation, especially for a wide field survey. 
In the CSST imaging simulator, spectra are required for all sources. However, given the observational depth and the vast number of stars, storing pre-computed spectra for the entire field would be prohibitively demanding in terms of storage. To address this, we adopt a strategy in which only the stellar catalogue is stored as input, while spectra are generated on the fly based on stellar parameters and apparent magnitudes. This requires the effective temperature $T_{\mathrm{eff}}$, the surface gravity $log\,\mathrm{g}$, the metallicity $z$, and magnitudes to be provided. Furthermore, to ensure astrometry function, the positions and velocities of all stars must also be included. For CSST, two widely used full-sky mock stellar catalogues are currently available and have been employed at different stages of development.

The first is the Milky Way stellar mock catalogue generated using the {\tt TRILEGAL} stellar population synthesis tool \citep{Chenyang2023SCPMA..6619511C}. 
This catalogue includes all classical components (the bulge, the disk and the halo). 
The luminosity function and the mass distribution are generally consistent with observational results \citep{Chenyang2023SCPMA..6619511C}. The {\tt TRILEGAL} catalogue provides detailed information of each star, including stellar parameters (effective temperature $T_{\mathrm{eff}}$, surface gravity $\mathrm{log\,}g$, metallicity $\mathrm{z}$, age $\tau$), the position $(\alpha,\delta,d)$, the velocity relative to the Sun $(v_{\rm U}, v_{\rm V}, v_{\rm W})$ and magnitudes in CSST photometric system ($NUV$, $u$, $g$, $r$, $i$, $z$, $y$), SDSS photometric system ($u_{\rm s}$, $g_{\rm s}$, $r_{\rm s}$, $i_{\rm s}$, $z_{\rm s}$) and PAN-STARRS photometric system ($g_{\rm p}$, $r_{\rm p}$, $i_{\rm p}$, $z_{\rm p}$, $y_{\rm p}$). What should be noticed is that only the stars with apparent magnitude brighter than 27.5 in CSST g-band are included. The dataset is accessible to all CSST simulator users via the Virtual Observatory platform \footnote{\url{ https://nadc.china-vo.org/data/data/csst-trilegal/f}}. Almost all spectral types  of stars are included while the interacting binary systems are not included for current simulations. 

The second model is {\tt Galaxia} \citep{Sharma2011ApJ...730....3S}, which is able to generate a synthetic survey of the Milky Way based on customizable parameters. Users can define the magnitude limit, sky coverage and output photometric system via a configuration file\footnote{\url{https://galaxia.sourceforge.net/Galaxia3pub.html}}. Similar with {\tt TRILEGAL}, the output catalogue includes the stellar magnitudes in the selected photometric system, as well as stellar parameters and chemical information. However, {\tt Galaxia} does not enforce a fixed magnitude cutoff, but instead adheres to the magnitude limits defined in the configuration file.
Another notable difference is that {\tt Galaxia} does not provide photometric magnitudes in the CSST filter system. Instead, we use the magnitudes available in the SDSS bands. The magnitudes in other photometric systems are recalculated from the spectra generated based on the stellar parameters.

Nonetheless, both {\tt TRILEGAL} and {\tt Galaxia} can support the requirements of CSST imaging simulations, as they provide the stellar parameters, ($T_{eff}$, $log\,g$, $z$) to reproduce the spectrum for each star, which is necessary for the simulator. At the current stage, we adopt {\tt Galaxia}, which is able to generate complete stellar samples down to the faint end for all CSST bands. As an example, the input star catalogue used for the current release is generated using {\tt Galaxia}. At this stage, crowded stellar systems, such as globular clusters and dwarf galaxies, are not included in the mock catalogue, but these larger systems will be incorporated in future versions. Fig.~\ref{fig:star_ra_dec} shows the stellar density distribution in equatorial coordinates for the catalogue covering the North Ecliptic Pole (NEP). Given that the NEP lies at a relatively low Galactic latitude ($b\sim29.81^\circ$), a distinct increase in stellar density appears in the lower-right region of the figure, corresponding to areas closer to the Galactic plane.

\begin{figure*}
    \centering 
    \includegraphics{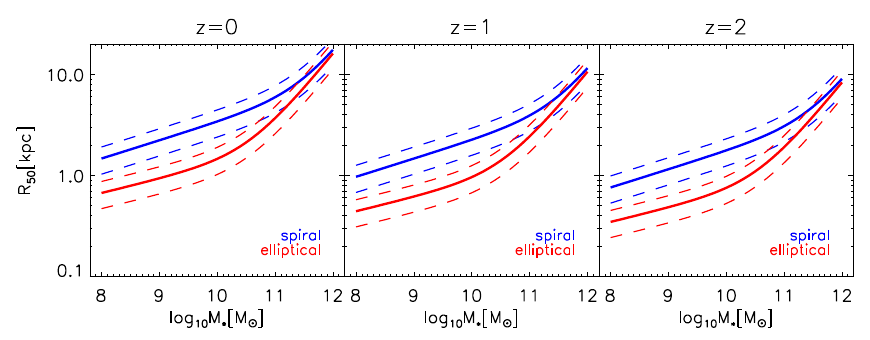}
    \caption{The mass-size relation of spiral galaxies and elliptical galaxies at $z=2$ to $z=0$ adopted in our mock galaxies catalogue. The dashed line indicates the $1\sigma$ scatter of the galaxy size, $R_{50}$. }
    \label{fig:mass-size}
\end{figure*}
\begin{figure*}
    \centering
    \includegraphics[width=\textwidth]{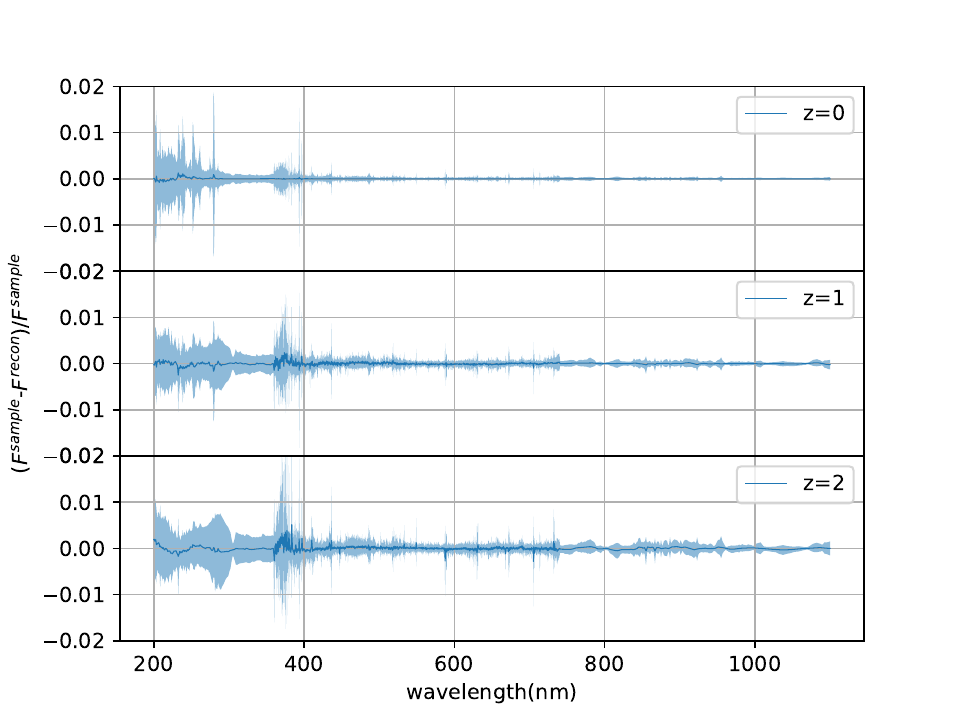}
    \caption{The relative deviation between the recovered SEDs and the SEDs of the original output of 10,000 galaxies at $z=0,1,2$. }
    \label{fig:pca}
\end{figure*}
\begin{figure*}
    \centering
    \includegraphics[width=0.75\textwidth]{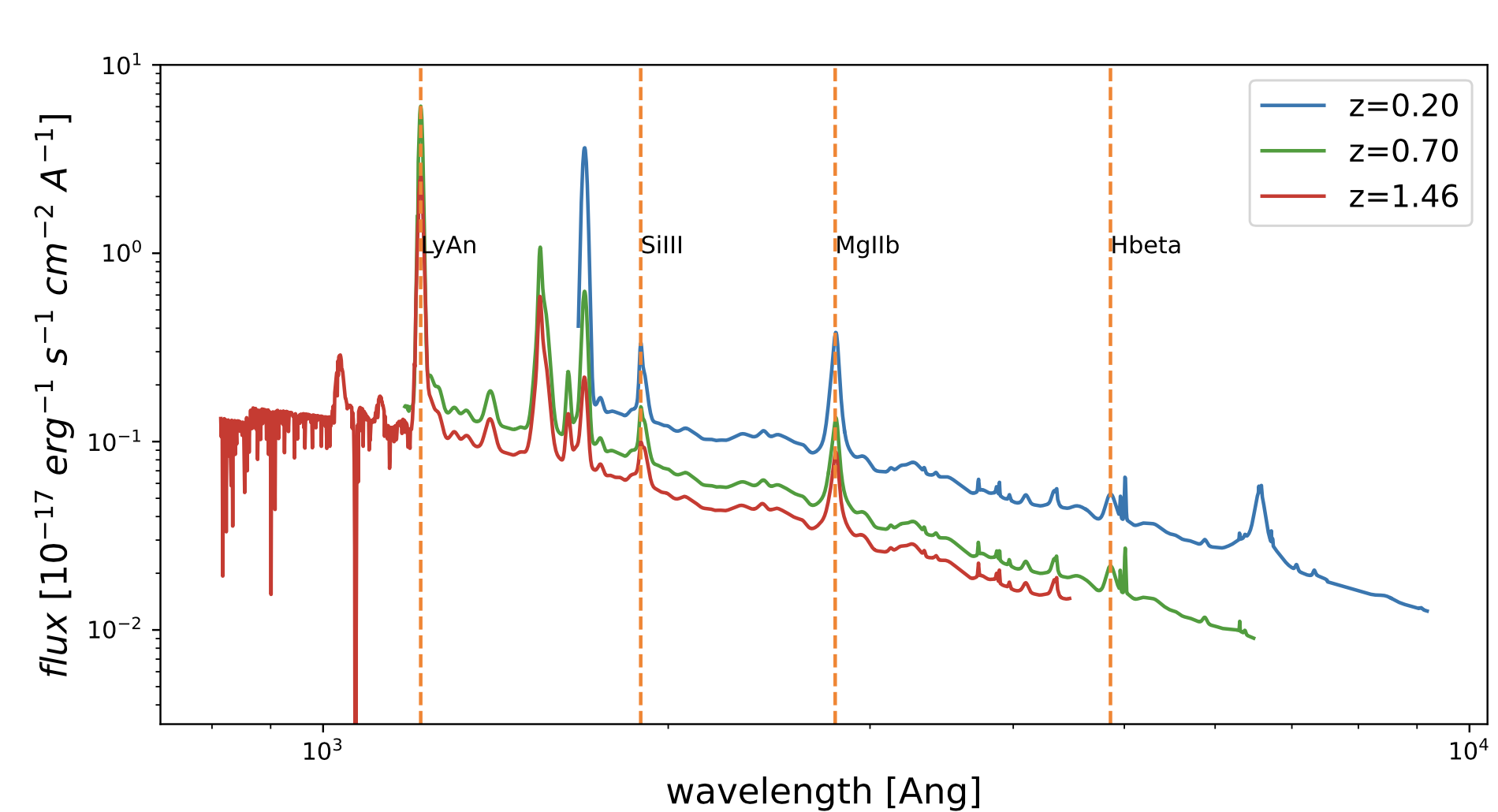}
    \caption{Three quasars' SEDs generated from the SIMQSO.}
    \label{fig:qso}
\end{figure*}

\section{Galaxy Catalogue}
\subsection{The Cosmological Dark Matter simulation}
\label{sect:dm simulation}
The `JiuTian' simulations are a series of $\Lambda$CDM cosmological N-body simulations with different simulation box sizes and resolutions, specifically designed for CSST \citep{Jiutian_2025arXiv250321368H}. The JT1G is one of the `JiuTian' simulation with both large volume and high resolution. It features $1 h^{-1} \rm {Gpc}$ box size and  $6144^3$ particles, which is 8 times larger volume and almost 27 times more particles than Millennium simulation (MS, \citealt{Springel+2005Natur_Millennium}). The corresponding dark matter particle mass is $3.72\times 10^8 h^{-1}\rm{M_{\odot}}$. For the purpose of weak lensing studies, JT1G stores 128 snapshots from redshift 127 to 0, with an average time interval of approximately 100 Myr. The adpoted cosmological parameters are based on Planck2020 results \citep{Aghanim+2020A&A_Planck}: $\sigma_8=0.8102$, $H_0=67.66\ {\rm km s^{-1} Mpc^{-1}}$, $\Omega_\Lambda=0.6889$, $\Omega_{\rm m}=0.3111$, $\Omega_{\rm b}=0.0490 ~(f_{\rm b}=0.1575)$. 

\subsection{Halo catalogue}
\label{sect:Halo}
The JT1G was executed using a lean version of the traditional Tree-PM code {\tt Gadget-3} \citep{2012MNRAS.426.2046A}. A significant portion of the typical post-processing work was performed on the fly, including group finding via the Friends-of-Friends algorithm \citep{Davis+1985ApJ_fof} and the extraction of bound substructures within each FoF group using the SUBFIND algorithm \citep{Springel+2001MNRAS_subfind}. 

The construction of merger trees follows a two-step approach. First, descendant links are established by tracking groups across adjacent snapshots. Once all unique descendant subhaloes are identified, we link them across the full snapshot sequence to construct complete merger trees. This is done by taking a subhalo at $z = 0$ and linking all subhaloes with descendant pointers to this halo, then repeating with all of those subhaloes, and so on, until no more subhaloes can be joined. Galaxies are then generated within these dark matter haloes using a galaxy formation model. Fig.~\ref{fig:lightcone} provides an illustrative example of our mock galaxy light-cone, where a pixel sky-field of {\tt HEALPix}\footnote{\url{healpix.jpl.nasa.gov}} with $N_{\rm side} = 64$ is displayed, and one can identify the large-scale structure of the universe at a given redshift snapshot.

\subsection{Semi-analytical modeling}
Galaxies are generated from halo merger trees using a semi-analytical galaxy formation model. In SAMs, galaxy populations are assigned to dark matter haloes based on simplified prescriptions for a wide range of physical processes, including reionization, hot gas cooling and cold gas infall, star formation and metal production, SN feedback, hot gas stripping and tidal disruption in satellites, galaxy mergers, bulge formation, black hole growth, and AGN feedback. A brief overview of these processes is provided below.

In most SAMs, each dark matter halo is assigned a cosmic abundance of baryons. As the halo grows, a significant fraction of baryons is accreted as primordial diffuse gas. This gas is shock-heated and then either cools rapidly onto the central galaxy's disk or is added to a quasi-static hot atmosphere, which accretes at a slower rate via cooling flows. The cold gas disk fuels star formation, and when some stars die, they release energy, mass, and heavy elements into the surrounding medium. This energy reheats the cold disk gas, injecting it into the hot atmosphere, while the hot atmosphere itself may be ejected into an external reservoir, only to be reincorporated much later.

Once a satellite crosses its host's virial radius, several environmental processes come into play. Tidal forces can strip both hot and cold gas and stars, while ram pressure stripping can remove gas. These processes gradually suppress star formation, especially in satellites orbiting more massive systems. As dark matter subhaloes merge, their associated galaxies merge as well, albeit with some delay. Once a subhalo is completely disrupted, its galaxy spirals into the central galaxy and merges after a dynamical friction time, forming a bulge and triggering a burst of star formation. Bulges can also form through a long-term process whenever the disk becomes dynamically unstable.

Black holes are believed to grow primarily through the accretion of cold gas during mergers, but also through static accretion of the hot atmosphere, releasing energy that can counteract cooling flows. This form of feedback eventually suppresses star formation in the most massive haloes. Finally, the light emitted by stellar populations of different ages is calculated using a population synthesis model and corrected for dust extinction.

In this project, we adopt the semi-analytical model developed by \cite{Luo+2016MNRAS}, which is based on the \cite{Guo+2013MNRAS} and \cite{Fu+2013MNRAS} model, a version of the Munich Semi-analytical model called L-Galaxies. \cite{Luo+2016MNRAS} improved the prescription for low-mass galaxies, especially satellite galaxies, by including additional physics of cold gas stripping and an analytical modeling of orphan galaxies. For more details, we refer readers to \cite{Luo+2016MNRAS}, \cite{Guo+2011MNRAS, Guo+2013MNRAS} and \cite{Fu+2012MNRAS, Fu+2013MNRAS}.

We introduce two major modifications for generating mock galaxy images. The first involves the galaxy size modeling. Similar to most SAMs, \cite{Luo+2016MNRAS} model assumes that each galaxy consists of two stellar components, an exponential disk and a ${\rm r}^{1/4}$-law bulge. The disk size is proportional to the ratio of specific angular momentum to maximum circular velocity. The growth of the bulge is mainly driven by mergers, with the bulge size being determined through energy conservation and the virial theorem for both minor and major mergers \citep{Guo+2011MNRAS}. These generated galaxy sizes are in roughly agreement with SDSS mass-size relation, but showing larger scatters for low mass galaxies. Accurately modeling the redshift evolution of the mass-size relation is also challenging. Therefore, we use the mass-size relation fitting formula for SDSS galaxies from \cite{Zhang+2019RAA} and consider a certain redshift evolution to describe the size of a given galaxy:
\begin{equation}
    R_{50}=\gamma(\frac{M_*}{M_0})^{\alpha}(1+\frac{M_*}{M_0})^{(\beta-\alpha)}(1+z)^\epsilon 
\label{eq:mass-size}    
\end{equation}
where $R_{50}$ is the half mass radius of the galaxy, $M_*$ is stellar mass and $z$ is redshift of the galaxy. For spiral galaxies,  $\alpha = 0.18$, $\beta=0.78$, $\gamma=6.34$, $M_0=1.57\times10^{11}\rm h^{-2}\rm M_{\odot}$, $\epsilon=-0.6$, for elliptical galaxies, $\alpha = 0.14$, $\beta=0.71$, $\gamma=1.53$, $M_0=1.72\times10^{11}\rm h^{-2}\rm M_{\odot}$, $\epsilon=-0.6$. Each galaxy is generated a Gaussian dispersion with $\sigma =0.3$. In Fig.~\ref{fig:mass-size}, we show the mass-size relation of spiral galaxies and elliptical galaxies from  $z=2$ to $z=0$ using the Eq.~\ref{eq:mass-size}. 

Another improvement is the approach used to generate the SED for each galaxy. Typically, in SAMs, the SED of a galaxy is generated by applying a stellar population synthesis (SPS) procedure, such as \cite{Bruzual+2003MNRAS}, with star formation history and geometry parameters of galaxy. To optimize calculation speed and storage, we employ the {\tt STARDUSTER} \citep{Qiu+2022ApJ} model to generate the high-resolution SEDs, which are then compressed using the principal component analysis (PCA). {\tt STARDUSTER} is a supervised deep-learning model trained using SKIRT radiative transfer simulation \citep{Camps+2015A&C_SKIRT} which can represent features of dust attenuation and emission features in galaxies. We input the star formation history, metallicity, size, and inclination angle of each galaxy from our mock galaxy catalogue into {\tt STARDUSTER}. We employ interpolation to resample the galaxy's rest-frame SED at uniform intervals of 0.29 nm across the 60-1100 nm wavelength range to meet the requirements for both PCA and slitless spectroscopy simulations. With a small sample of SEDs, we generate 20 principal components (PCs) and decompose each galaxy’s SED using these 20 PCs. Therefore, the SED of each galaxy is represented by 20 coefficients ($w_i$), from which the full SED can be reconstructed using $F(\lambda)=\sum w_i {\rm PC}_i(\lambda)$. To assess the accuracy of the SED reconstruction, we randomly select about 10,000 galaxies at redshifts 0,1,2. As shown in Fig.~\ref{fig:pca}, the comparison between the reconstructed and original SEDs shows that the relative deviation is below $2\%$, indicating a high level of fidelity in the PCA-based compression.

\begin{figure*}
    \centering
    \includegraphics[width=\textwidth]{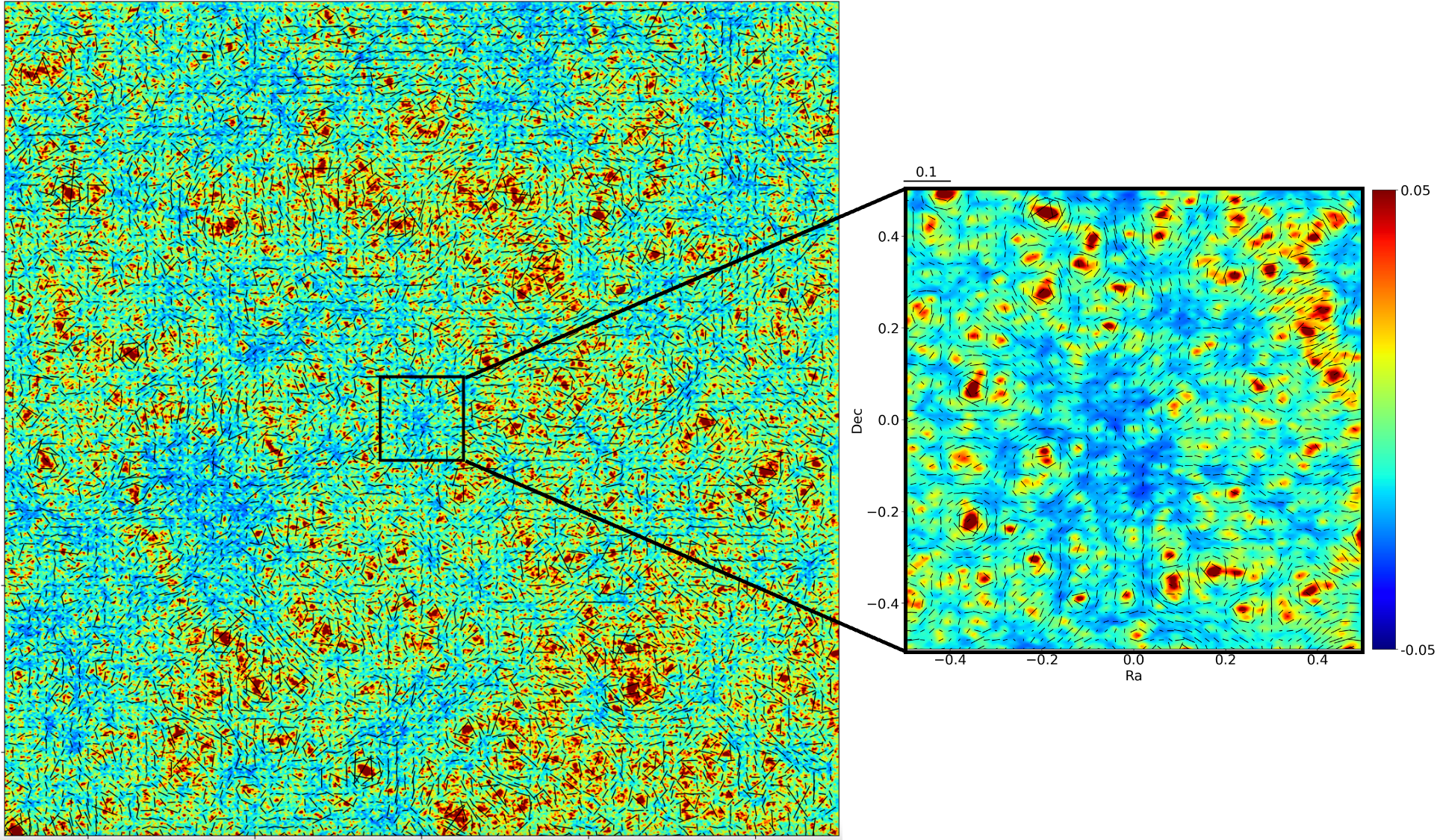}
    \caption{The convergence (colour map) and shear field (sticks) for sources at $z_s=1$ over a patch of $10\times10 \deg^2$ (left) and in a zoom-in of the central $1\times1 \deg^2$ area (right). }
    \label{fig_lens0}
\end{figure*}

%\subsection{Quasar catalogue}
The quasar catalogue is also generated based on galaxy catalogue. We use {\tt SIMQSO}\footnote{\url{https://simqso.readthedocs.io/en/latest}} \citep{McGreer+2021ascl.soft06008M} to generate quasar SEDs. {\tt SIMQSO} is a toolset that generates mock quasar spectra using a broken power-law continuum model along with Gaussian emission line templates. We calculate the total number of quasars in a target sky area using quasar luminosity functions. Specifically, we integrate the quasar luminosity functions from \citep{Ross+2013ApJ} over redshift bins of $\Delta z=0.1$ ranging from $z=0.1$ to $z=3.5$ to determine the total number of quasars within the target sky area. We then generate the SEDs of individual quasars based on their luminosities. Finally, we assign these quasars to galaxies in the same redshift bins and match their luminosities to the AGN accretion rates of the corresponding galaxies from the galaxy catalogue. In Fig.~\ref{fig:qso}, we present examples of quasar SEDs generated using {\tt SIMQSO}. 

The quasar catalogs are generated alongside the galaxy catalogue to enable joint simulations. We use {\tt SIMQSO} to create the mock quasar catalogs through the following steps: First, quasars are randomly sampled according to the observed SDSS quasar luminosity functions (Ross et al. 2013), producing distributions in redshift and absolute magnitude (M1450) that are consistent with observations at a redshift range of [0.1 to 2.1]. This ensures the simulated quasars closely match the observed quasar populations. Second, we generate individual quasar SEDs using {\tt SIMQSO}’s default configuration, specifically adopting a broken power-law continuum model combined with Gaussian emission line templates, based on the luminosities (M1450) determined in the first step. Finally, these simulated quasars are assigned to galaxies within corresponding redshift bins, matching their luminosities to the AGN accretion rates derived from the galaxy catalog.

\section{Weak lensing maps}
To produce weak lensing maps from a cosmological simulation, we employ a multi-plane algorithm on spherical geometry. Below, we briefly summarize the key procedures of our weak lensing simulation and more details can be found in \cite{Wei+2018ApJ}.

We first cut the simulation box of JT1G into a collection of small cubic boxes with $100 h^{-1} {\rm Mpc}$ on each side and assemble the cell boxes to cover a past light-cone from $z=0$ to $z_{\rm max} \sim 3.5$. Next, the light-cone is decomposed into a series of spherical shells of a given width, $50\ h^{-1}{\rm Mpc}$, around the observer, and the dark matter particles are projected onto their corresponding shells, pixelized
by the {\tt HEALPix} tessellation \citep{Gorski+2005ApJ_Healpix}. In this work,
the {\tt HEALPix} resolution is set to $N_{\rm side} = 8192$, which corresponds to
an angular resolution of ${\rm d}\theta \sim 0.43\ {\rm arcmin}$ \footnote{The
pixel size of {\tt HEALPix} cells can be calculated by
${\rm d}\theta = \left( 4\pi/12N_{\rm side}^{2}\right)^{1/2}$ for a given
$N_{\rm side}$.}. The projected surface mass densities are computed with an Epanechnikov kernel for
each shell and the smoothing length for each particle is set to a few N-body softening lengths. Finally we obtain a surface matter overdensity $\Sigma^{(n)}$ on the $n$-th shell by
\begin{equation}
  \Sigma^{(n)}(\bm{\theta}^{(n)}) = \int^{\chi_{n+1/2}}_{\chi_{n-1/2}}
                                    {\rm d}\chi' \delta(r(\chi')\bm{\theta}^{(n)}, \chi').
\end{equation}
where $\delta = \rho/\bar{\rho} - 1$ is the mass overdensity. The convergence field is $\kappa^{(n)}(\bm{\theta}^{(n)})  = W^{(n)}\Sigma^{(n)}(\bm{\theta}^{(n)})$, where lensing kernel function $W^{(n)}$ is defined as
\begin{equation}
  W^{(n)} = \frac{3}{2}\Omega_{\rm m} \left( \frac{H_{0}}{c} \right)^{2}
            \frac{r(\chi_{n})}{a(\chi_{n})}.
\end{equation}
Using the Poisson equation, the lensing potential $\phi^{(n)}$ and deflection field $\bm{\alpha}^{(n)}=\nabla \phi^{(n)}$ can be derived from the mass density of a shell in spherical harmonic space
\begin{equation}
  \phi_{\ell m}^{(n)} = -\frac{2}{\ell (\ell+1)} \kappa_{\ell m}^{(n)}
  %\nabla^2 \phi^{(n)} = 2\kappa^{(n)}
\end{equation}
where $\kappa_{\ell m}^{(n)}$ are the spherical harmonic coefficients of
the convergence $\kappa^{(n)}$.

After obtaining the deflection field on each lens plane, we can perform ray-tracing simulation between different lens planes. Light rays are initialized at the center of each {\tt HEALPix} cell in the first shell and are then propagated to the next lens plane according to $\bm{x}^{(n+1)} = \mathcal{R}(\bm{n}^{(n)}\times\bm{\alpha}^{(n)}, \lVert\bm{\alpha}^{(n)}\rVert)  \bm{x}^{(n)}$, where $\bm{n}$ denotes the radial direction of light rays on the $n$-th shell and the rotation matrix $\mathcal{R}$ makes a rotation with an angle $\lVert\bm{\alpha}\rVert$ to the light rays. The lensing distortion matrix $\mathcal{A}^{(n+1)}$ can be calculated at the position of light rays by
\begin{equation}
\begin{aligned}
  \mathcal{A}^{(n+1)}_{ij}
   = & \left(1-\frac{D^{n}_{0}}{D^{n+1}_{0}}\frac{D^{n+1}_{n-1}}{D^{n}_{n-1}} \right) \mathcal{A}^{(n-1)}_{ij} \\
     & + \frac{D^{n}_{0}}{D^{n+1}_{0}}\frac{D^{n+1}_{n-1}}{D^{n}_{n-1}} \mathcal{A}^{(n)}_{ij}
   - \frac{D^{n+1}_{n}}{D^{n+1}_{0}} \mathcal{U}^{(n)}_{ik}\mathcal{A}^{(n)}_{ij},
\end{aligned}
\end{equation}
where $D^{n+1}_{n}$ is the comoving angular diameter distance,
$D^{n+1}_{n} \equiv r(\chi_{n+1}-\chi_{n})$. $\mathcal{U}^{(n)}_{ij}$
denotes the deformation matrix in $n$-th lens plane, which can be
calculated as $\mathcal{U}^{(n)}_{ij} = \phi^{(n)}_{,ij}$. 

\begin{figure}
    \centering
    \includegraphics[width=0.45\textwidth]{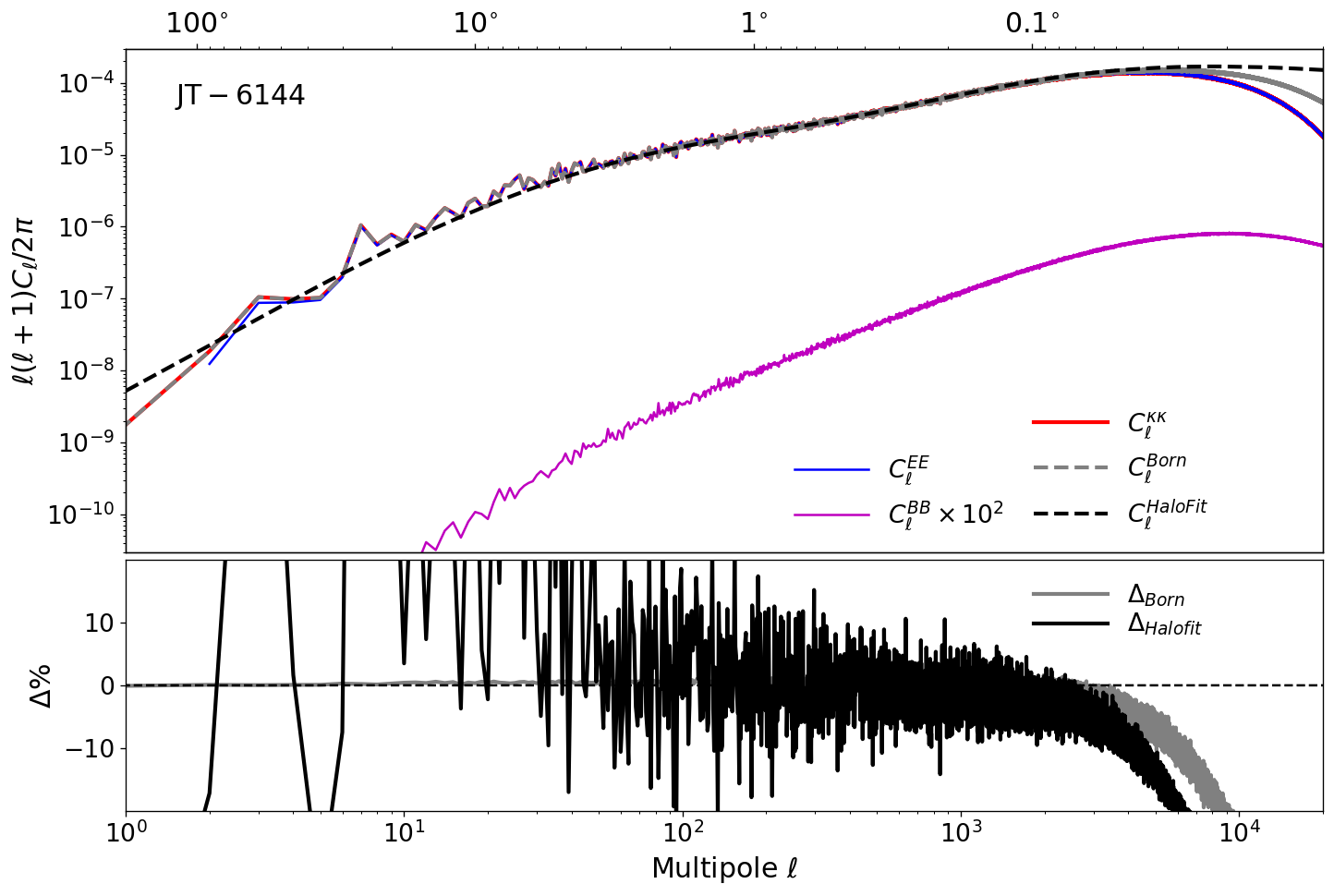}
    \caption{Angular power spectra of the convergence field at $z_s = 1$. Plots show measurements in the simulation (red solid) compared against the predictions from Born approximation (gray dashed) and Halofit model (black dashed, Takahashi et al. 2012), respectively. The power spectra of the shear E-mode (blue) and B-mode (magenta) are also included for comparison. The lower panels show the relative errors between the simulation and theoretical predictions.}
    \label{fig_lens1}
\end{figure}
% \begin{figure}
%     \centering
%     \includegraphics[width=0.45\textwidth]{figures/fig_lens2.png}
%     \caption{The angular power spectrum of the convergence (red solid line), shear E-mode (blue) and B-mode (magenta), and rotation (cyan) for sources at $z_s = 1$. The measured convergence powers from the Born approximation (gray dashed line) and revised Halofit model predictions (black dashed line) are also shown for comparison. }
%     \label{fig_lens2}
% \end{figure}

\begin{figure}
    \centering
    \includegraphics[width=0.5\textwidth]{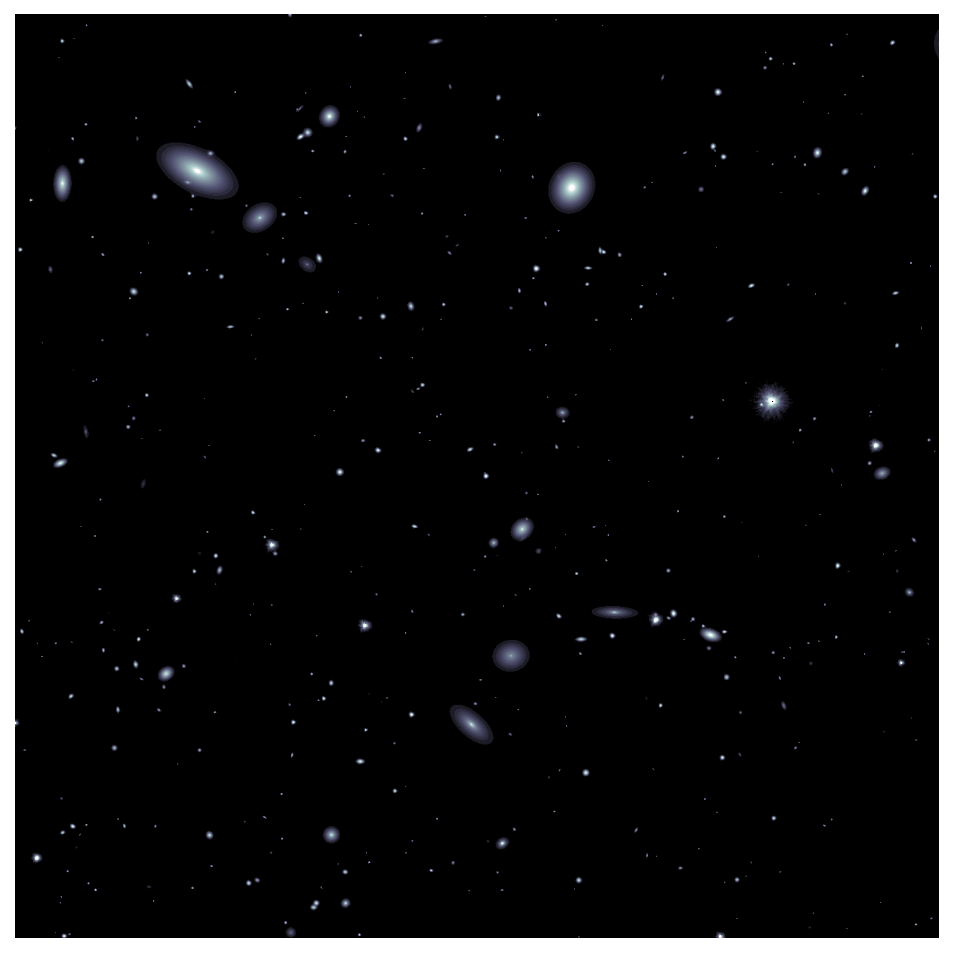}
    \caption{Illustration of the input mock catalogue within an sky field of $\sim 1.5^{\prime} \times 1.5^{\prime}$.}
    \label{fig_skymap}
\end{figure}

 \begin{figure*}
    \centering
    \includegraphics[width=0.49\textwidth]{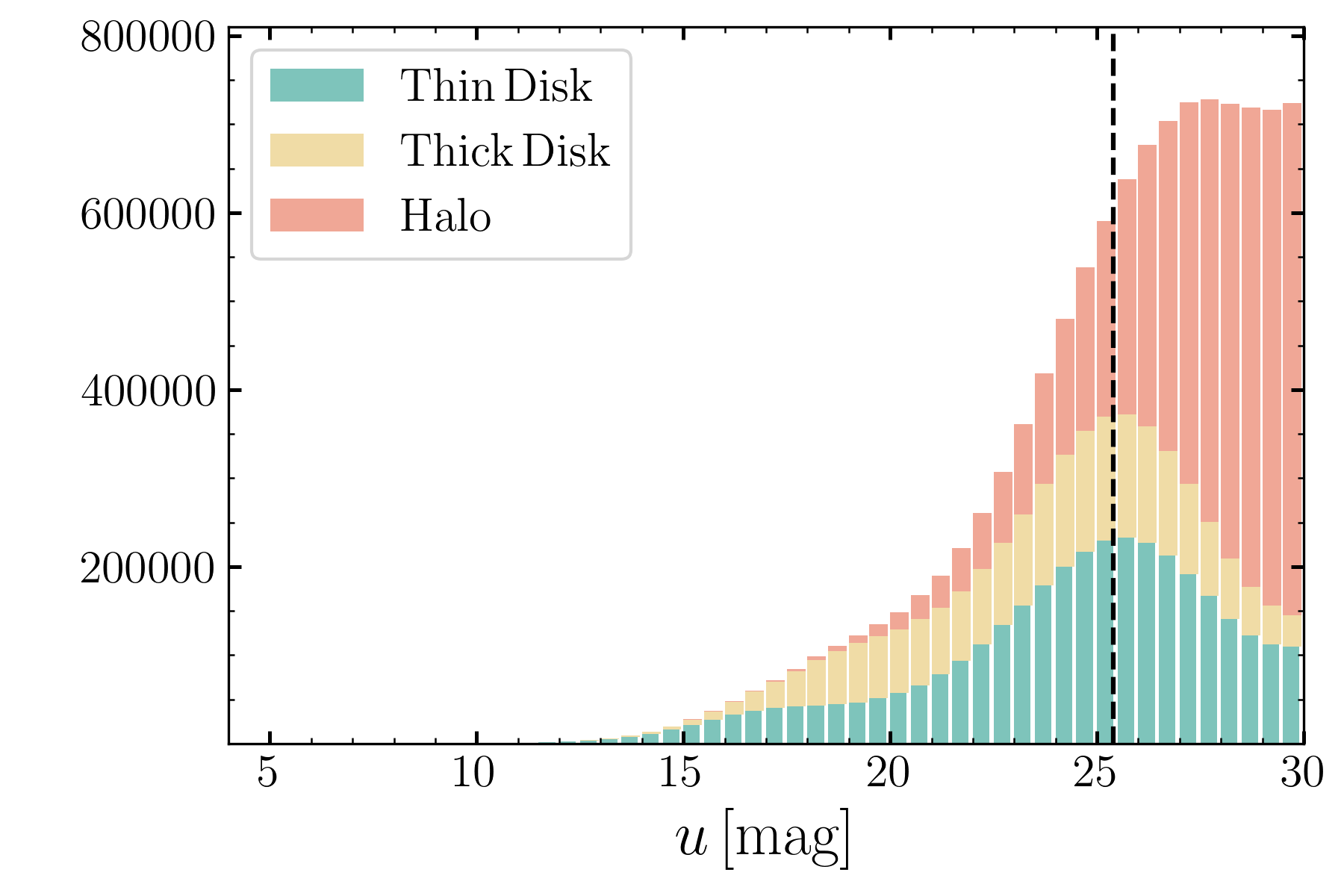}
    \includegraphics[width=0.49\textwidth]{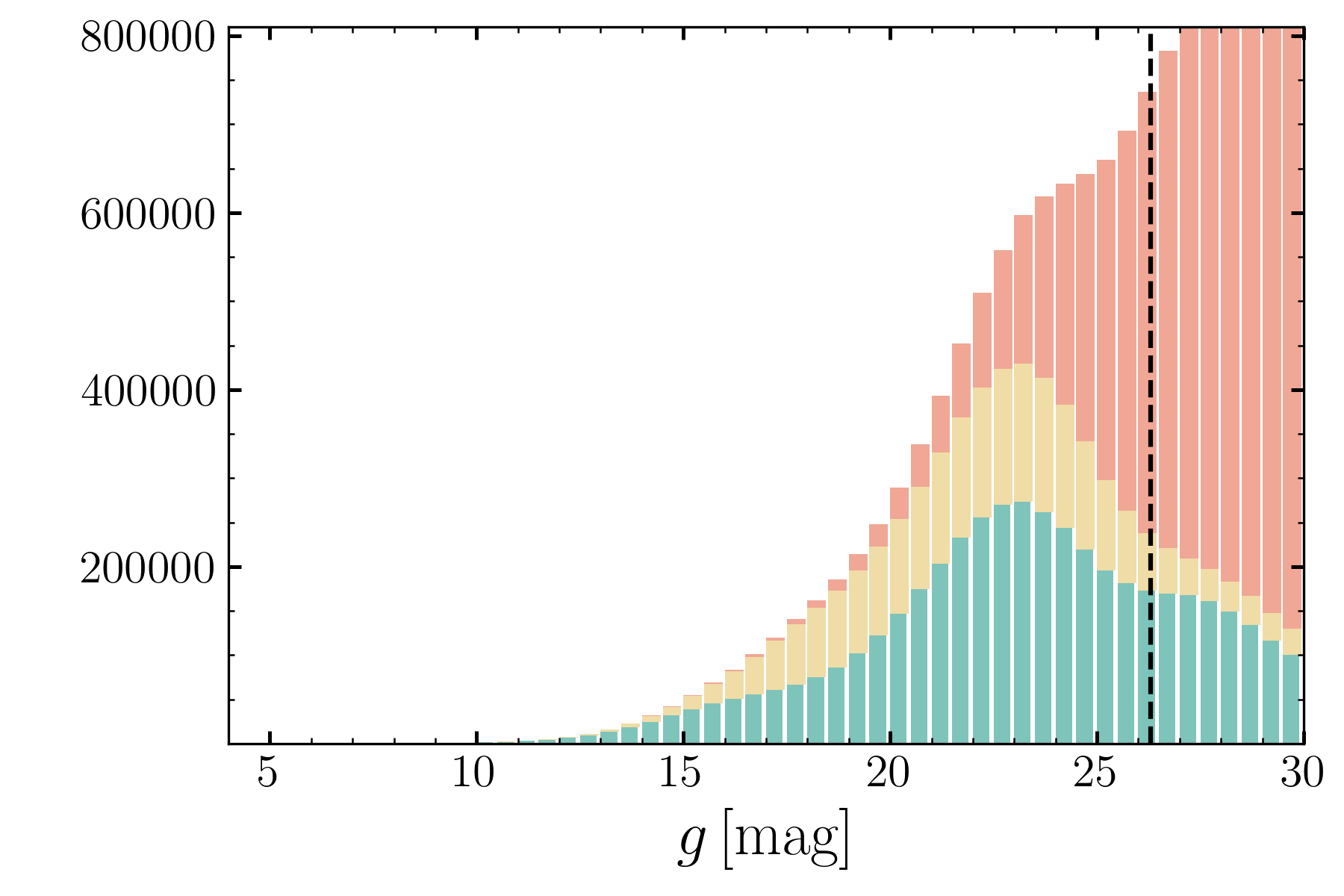} \\
    \includegraphics[width=0.49\textwidth]{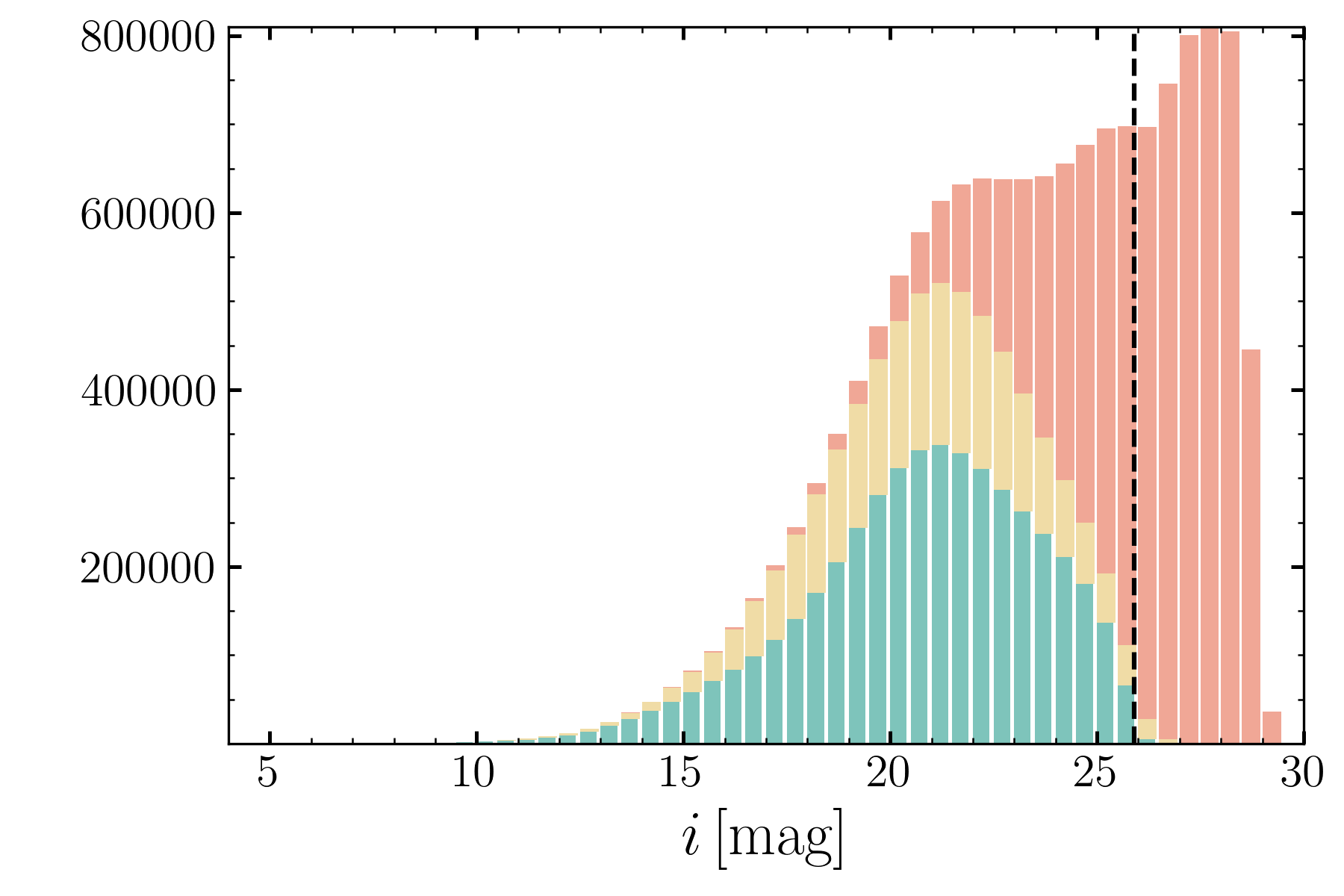}
    \includegraphics[width=0.49\textwidth]{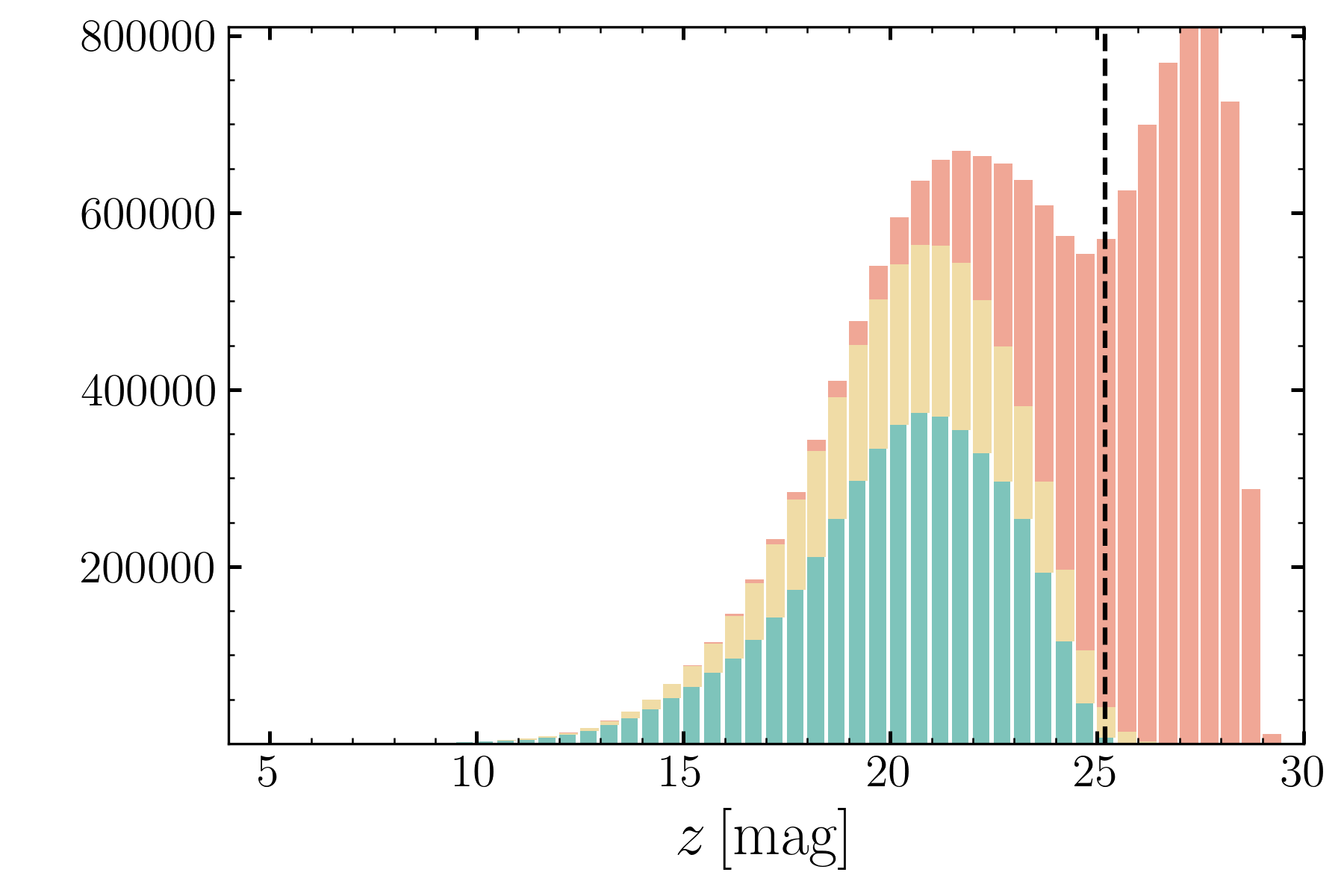}
    \caption{The luminosity functions in u, g, i, z band
    of all the stars and the stars from the thin
    disk, the thin+thick disk and the thin+thick+halo are represented by 
    different colors in the panels from top left to the bottom right. 
    The limit magnitude in each band is represented by the black dashed line. }
    \label{fig:star_hist_mag}
\end{figure*}

Fig.~\ref{fig_lens0} displays a patch of $10 \times 10 \deg^2$ from the all-sky map of the convergence field for sources at $z_{\rm s} = 1.0$. In the figure, the short sticks are overlaped to indicate the lensing shear in the field. As expected, galaxies are stretched tangentially around convergence peaks by the gravitational lensing. As a basic validation of the lensing maps, Fig.~\ref{fig_lens1} shows the measurement of the convergence angular power spectrum in the simulation (solid) compared with theory predictions from Born approximation (gray dashed) and Halofit model (black dashed, \citealt{Takahashi+2012ApJ_Halofit}), for source redshifts at $z_{\rm s} = 1$. The measured convergence power from the ray-tracing simulation shows good agreement with theoretical predictions within weak lensing scales, with a relative error of less than 10\% at  $l \leq 4000$. For comparison, we also show the power of the shear B-mode (magenta) for sources at $z_{\rm s} = 1.0$, which is suppressed by more than four orders of magnitude relative to the E-mode (blue), indicating that numerical artifacts and residual systematics are effectively mitigated in the simulation.
% Fig.~\ref{fig_lens2} presents the angular power spectrum of the convergence (red solid line), shear E-mode (blue) and B-mode (magenta), and rotation (cyan)  with sources at $z_s = 1$. For comparison, the convergence power derived from the Born approximation (gray dashed line) and the revised Halofit model (black dashed line) are also included. In line with lensing theory, at high-$\ell$(small scales) the relation $C_{\rm EE} (\ell) = C_{\kappa\kappa} (\ell)$ holds in our full-sky weak lensing simulation and at low-$\ell$ (large scales) the difference between $C_{\rm EE} (\ell)$ and $C_{\kappa\kappa} (\ell)$ arises from the additional factor $(\ell-1)(\ell+2) \ell (\ell+1)$. Additionally, we also measure the power spectra of the B-mode from lensing simulation and find that B-mode power is effectively suppressed by more than four orders of magnitude compared to the E-mode. The accuracy of the shear map is primarily limited by the smoothing length at small scales. 

Finally, as an example, we visualize our synthetic star and galaxy catalogue in Fig.~\ref{fig_skymap}, where the CSST optical PSFs are used to generate the image. Point sources (stars or quasars) appear as PSFs and galaxies are modeled by a Sersic profile with their morphological parameters. The weak lensing effect is applied to shear the galaxy shapes. In the next section, we will validate the statistic properties of the mock star and galaxy catalogue, respectively.

 \begin{figure*}
    \centering
    \includegraphics[width=0.45\textwidth]{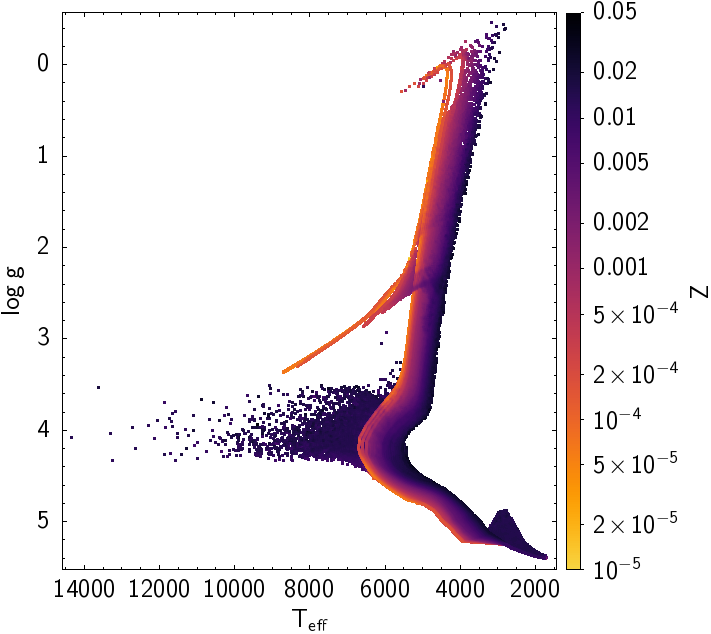} 
    \includegraphics[width=0.45\textwidth]{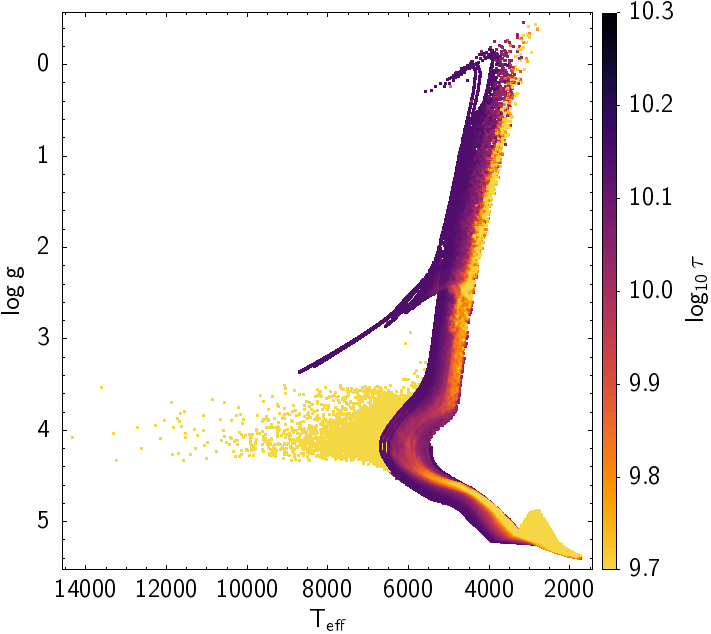}
    \caption{The Kiel diagram of the stars are shown, with color coded by the metallicity and age in the left and right panels, respectively.}
    \label{fig:star_CMD}
\end{figure*}

\begin{figure*}
    \centering 
    \includegraphics[width=0.9\textwidth]{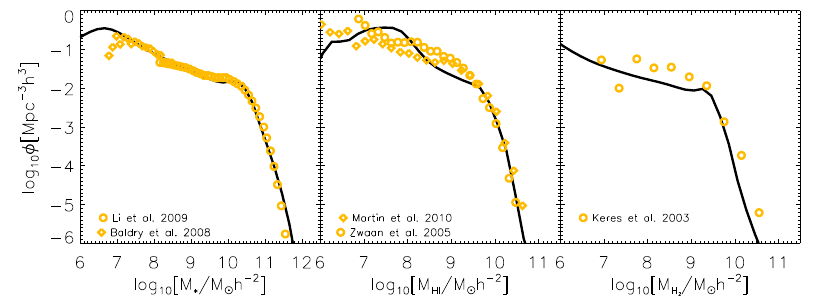}
    \caption{The mass functions of stellar mass, $\rm{H_I}$ and $\rm{H_2}$ at $z=0$. Yellow symbols are the observation data from \cite{Li+2009MNRAS, Baldry+2008MNRAS, Martin+2010ApJ, Zwaan+2005MNRAS, Keres+2003ApJ}. }
    \label{fig:mf}
\end{figure*}

\begin{figure*}
    \centering 
    \includegraphics[width=0.85\textwidth]{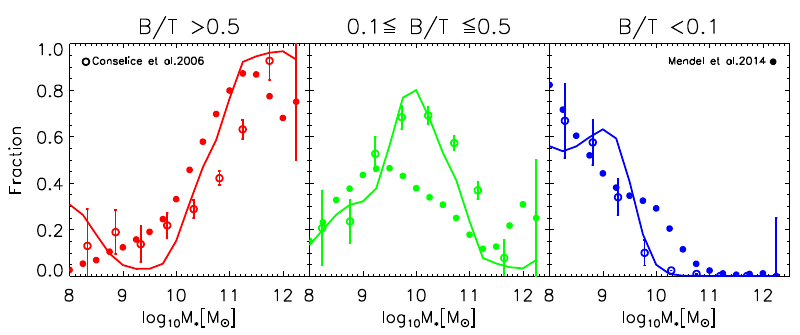}
    \caption{The galaxy population distributions with elliptical galaxies (red), spiral galaxies (green) and pure disk galaxies (blue) at $z=0$. The open cycle and dot are from \cite{Conselice+2006MNRAS} and \cite{Mendel+2014ApJS}.}
    \label{fig:fmorph}
\end{figure*}
\begin{figure}
    \centering 
    \includegraphics[width=0.45\textwidth]{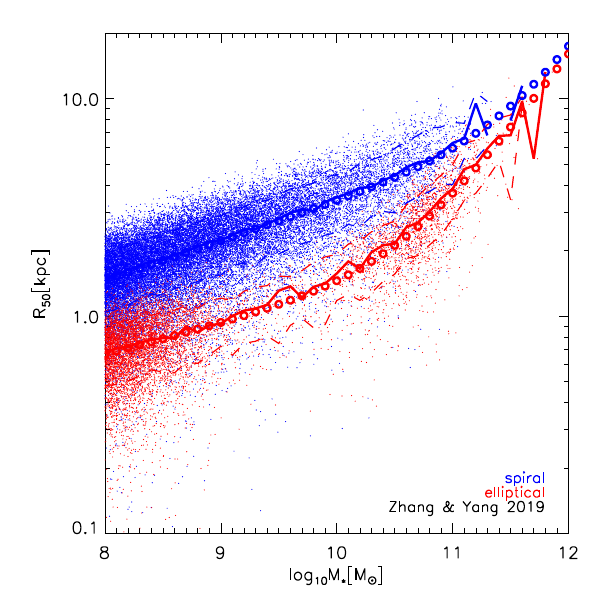}
       \caption{The mass-size relation at $z=0$, separated for spiral galaxy (blue dots) and elliptical galaxy (red dots). Solid lines with dashed lines are shown the median $\rm{R_{50}}$ and 1 $\sigma$ area. The open dots are from the \cite{Zhang+2019RAA}. }
    \label{fig:mass-size-0}
\end{figure}

\begin{figure*}
    \centering 
    \includegraphics[width=0.98\textwidth]{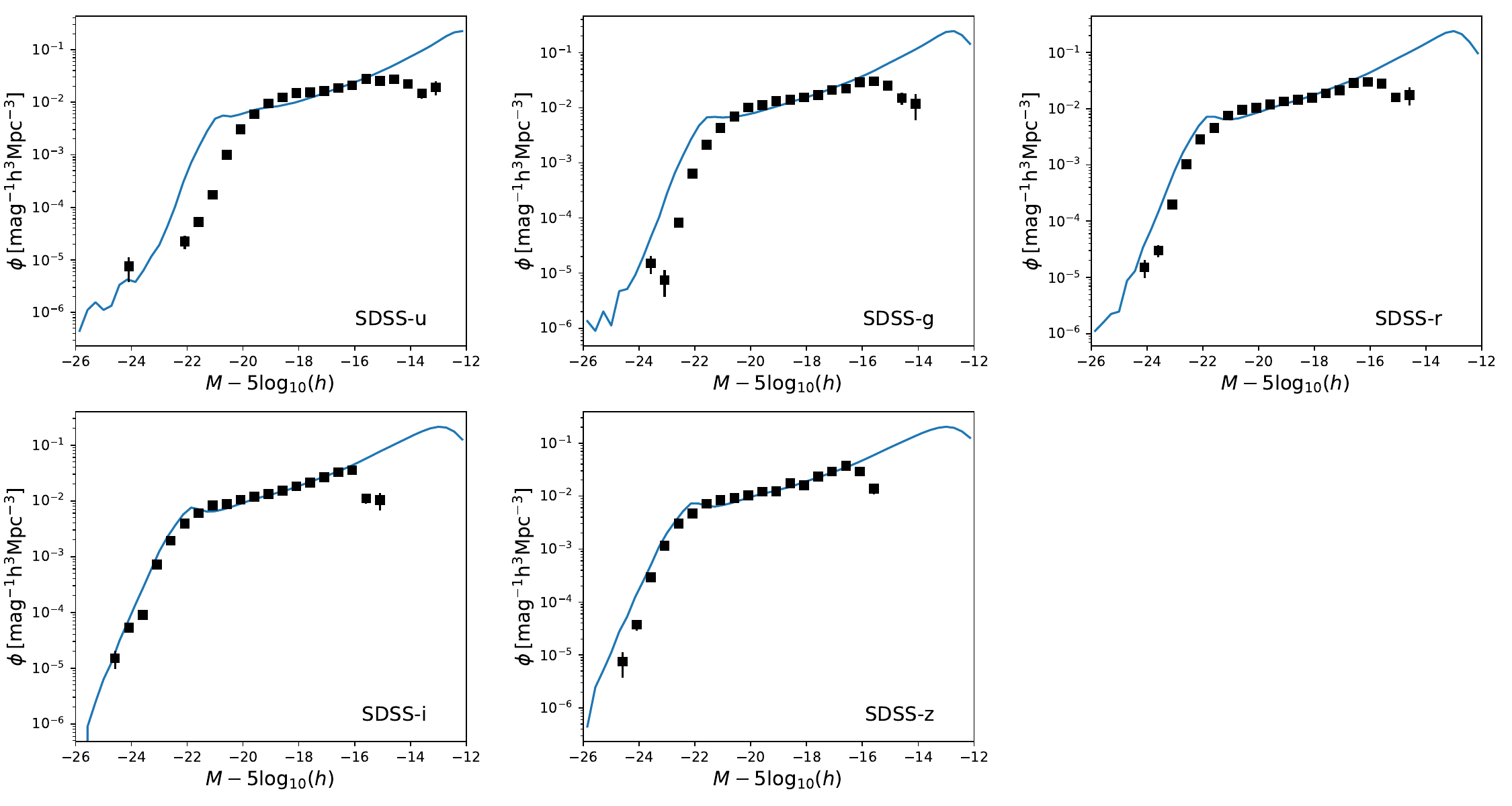}
    \caption{The luminosity function of SDSS $ugriz$ bands at $z=0$, the black symbol are the SDSS observation data from \cite{Driver+2012MNRAS}.}
    \label{fig:lf}
\end{figure*}

\section{Mock validation}
\label{sect:validation}
\subsection{Mock star properties}
During the selection of the star catalogue, we apply a magnitude cut to ensure the sample's completeness. Since {\tt Galaxia} does not provide data in the CSST photometric system, we adopt the SDSS system and keep all stars with an apparent magnitude of $g_{\mathrm{SDSS}}<30$. This threshold ensures completeness down to the limiting magnitudes across all seven CSST bands, as the color indices of different stellar types vary significantly. Fig.~\ref{fig:star_hist_mag} shows the luminosity distributions in the four bands, $u$, $g$, $i$, $z$. The vertical dashed lines indicate the CSST limiting magnitudes in the $u$, $g$, $i$, and $z$ bands, which are 25.4, 26.3, 25.9, and 25.2, respectively. The distributions are not always increasing exponentially versus the magnitude, because the contribution from the thin and thick disks are limited within a few kiloparsec. In contrast, the halo, being the most diffuse component, extends from a few parsecs to beyond 100 kpc. Consequently, its luminosity function closely follows an exponential distribution with magnitude.

Regarding the stellar populations, the mock Milky Way catalogue includes different populations relative to the components, i.e. the thin disk, the thick disk and the halo from young to old. Fig.~\ref{fig:star_CMD} shows the distributions of the stars in the Kiel diagram, with the left and right panels color-coded by metallicity and age, respectively. This catalogue covers a wide range of stellar populations, from metal-poor stars with $\mathrm{Z} \sim 1e-6$ to metal-rich stars with $\mathrm{Z}\sim0.1$, from hot stars with $T_{\mathrm{eff}}\sim14000$ K to cool stars with $T_{\mathrm{eff}}\sim2000$ K. What should be noticed is that certain special stellar types, such as chemically enhanced stars, are not included in this dataset.

\subsection{Mock galaxy properties}
In general, several model parameters in SAMs can be adjusted to better match observations. In this work, we constrain our model using stellar and cold gas mass functions at $z=0$. Figure~\ref{fig:mf} presents a comparison of our galaxy catalogue's stellar mass function (left panel), $\rm{H_I}$ mass function (middle panel) and $\rm{H_2}$ mass function (right panel) with observational data (yellow symbols) from \citealt{Li+2009MNRAS, Baldry+2008MNRAS, Martin+2010ApJ, Zwaan+2005MNRAS, Keres+2003ApJ} at $z=0$.

Our mock galaxy catalogue includes key parameters such as redshift, position, velocity, halo mass, stellar mass (separated into disk and bulge mass), galaxy type (central or satellite), size, and SED represented by 20 coefficients. In Appendix~\ref{appendixA}, we present statistical analyses, including the conditional stellar mass function, the evolution of the stellar mass function, and the stellar mass–halo mass relation from $\rm z=2$ to $0$, comparing them with the results derived from observational data. Our mock catalogue can reproduce the fundamental stellar mass and halo mass relation of various galaxy populations.

The galaxy size and morphology are two important properties to generate galaxy mock images. Normally we use B/T which is the ratio of bulge mass and total stellar mass to define the galaxy morphology \citep{2009MNRAS.396.1972P}. In Fig.~\ref{fig:fmorph} we show three galaxy population distributions with different morphology at $z=0$. Here, we define elliptical galaxy with $\rm{B/T} > 0.5$, spiral galaxy with $0.1 \leq \rm{B/T} \leq 0.5$ and pure disk galaxy with $\rm{B/T} < 0.1$. Our results can roughly reproduce the observation data \citep{Conselice+2006MNRAS, Mendel+2014ApJS}. We then assign an intrinsic shape to each galaxy using the mass distribution of its dark matter halo. For an early-type central galaxy, its major axis is assumed to align with that of the host dark matter halo. For a late-type central galaxy, its spin follows that of the halo, with the major axis determined by projecting the circular disk onto the sky. For satellite galaxies, we assign their shapes using a random orientation. A detailed description of this model can be found in \cite{Wei+2018ApJ}.

In Fig.~\ref{fig:mass-size-0}, we show the mass-size relation of our galaxy catalogue at $z=0$. We simply divide a small sample of our galaxy catalogue into two populations, spiral galaxy with  $\rm{B/T} \leq 0.5$ (blue dots), elliptical galaxy with $\rm{B/T} > 0.5$ (red dots). Solid lines represent the $\rm{R_{50}}$, while the corresponding dashed lines indicate the error with 1 $\sigma$. The open dots are from the \cite{Zhang+2019RAA}. Since we use the fitting formula from \cite{Zhang+2019RAA}, it can reproduce the relation of the observational data.

The magnitude of each galaxy is computed by convolving its SED with a given filter. In Fig.~\ref{fig:lf}, we present the luminosity functions of our mock galaxy catalogue in $u$, $g$, $r$, $i$, $z$ bands of SDSS at $z=0$, which closely match SDSS observational data (black symbols; \citealt{Driver+2012MNRAS}). The only exception is the u-band, where our catalogue contains a higher number of bright galaxies, likely due to the stellar population synthesis (SPS) model used.

\begin{figure}
    \centering
    \includegraphics[width=0.475\textwidth]{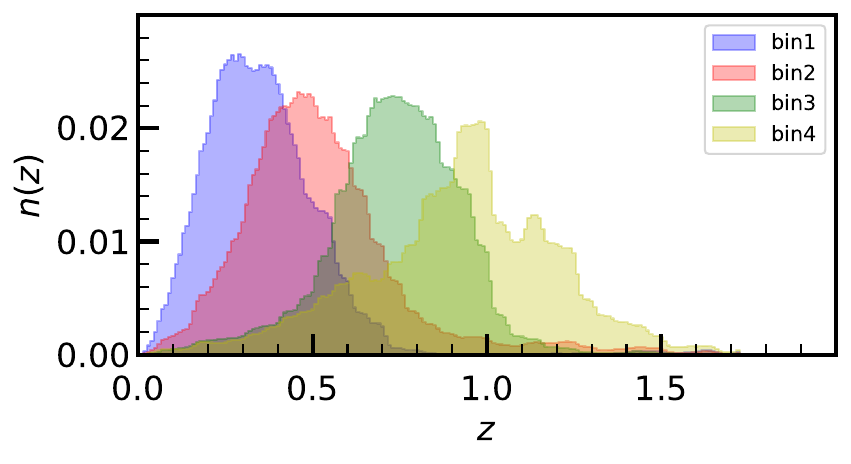}
    \caption{Redshift distributions of the source galaxies for tomographic analyses in our DES-Y3-like mock catalogue.}
    \label{fig_lens3}
\end{figure}

\begin{figure*}
    \centering
    \includegraphics[width=\textwidth]{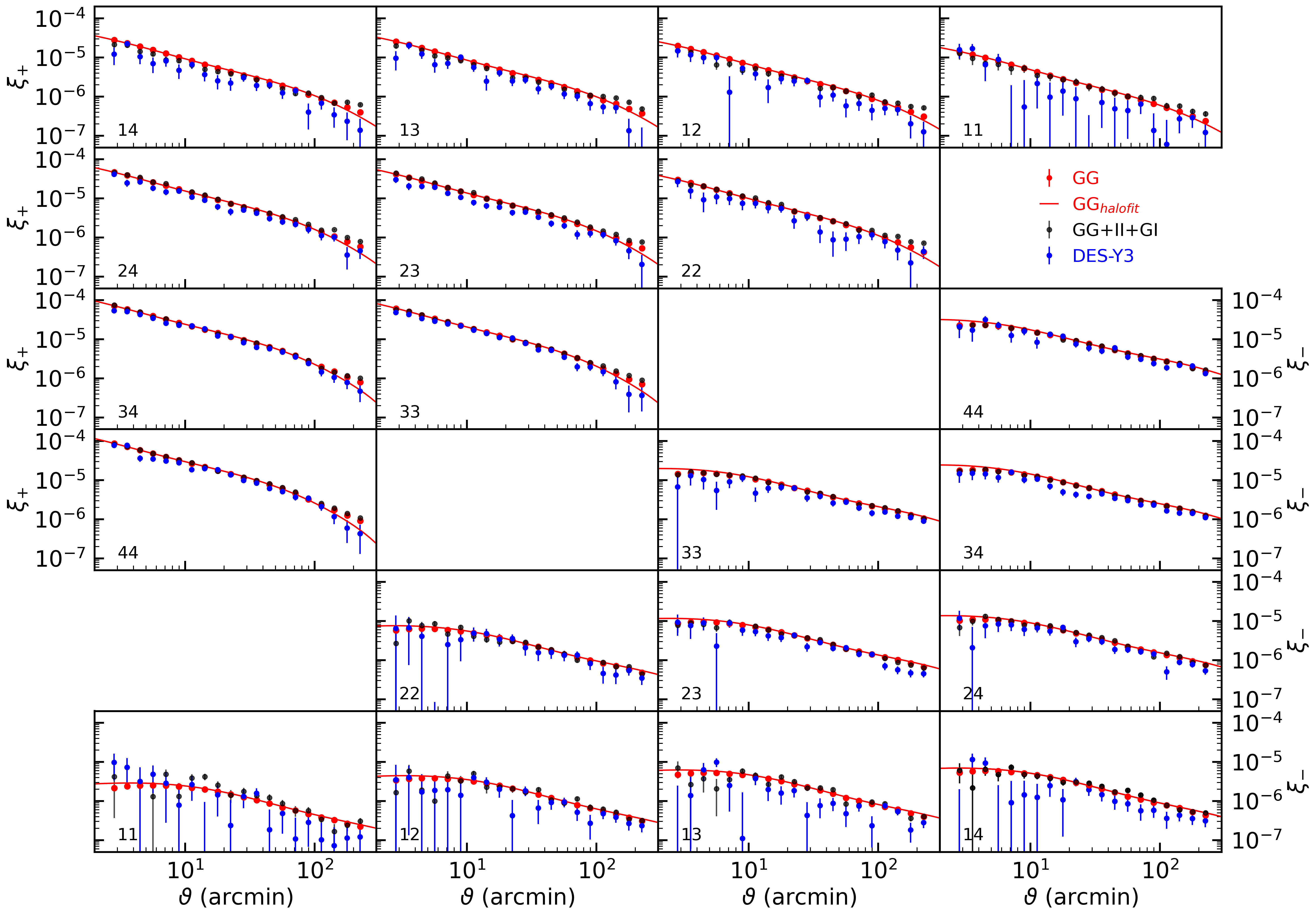}
    \caption{Tomographic measurements of $\xi_\pm$ from our mock catalogue at four different redshift bins. The red circles represent the shear correlations from the gravitational field (GG). The solid lines shows the theoretical predictions from the halofit model. The black circles are the total shear correlations (GG+II+GI), which can be directly compared with the observational results (blue circles) from DES-Y3.}
    \label{fig_lens4}
\end{figure*}

\subsection{Weak lensing 2-point statistics}
Weak lensing can result in small correlated distortions of galaxy shapes, which can be analyzed using various statistical methods \citep{Munshi+2011MNRAS, Shirasaki+2012ApJ, Fu+2014MNRAS_cfhtlens, Shan+2018MNRAS}. A standard statistic used in weak lensing analysis is the shear 2-point correlation function \citep{Bartelmann+2001PhR}. The shear 2-point correlation between galaxies at separation $\vartheta$ is usually defined as
\begin{equation}
  \begin{aligned}
  & \xi_{\rm tt}(\vartheta) = \frac{\sum_{i,j} w_{i}w_{j}\varepsilon_{t,i}\varepsilon_{t,j}}{\sum_{i,j}w_{i}w_{j}}
  \\
  & \xi_{\times\times}(\vartheta) = \frac{\sum_{i,j} w_{i}w_{j}\varepsilon_{\times,i}\varepsilon_{\times,j}}{\sum_{i,j}w_{i}w_{j}}
  \end{aligned}
\end{equation}
where $\varepsilon_{t}$ and $\varepsilon_{\times}$ are the tangential and cross-component of weak lensing shear, $w_{i}$ is a weight of the $i$-th galaxy. A more commonly used form is the linear combination of $\xi_{\rm tt}$ and $\xi_{\times\times}$, i.e. $\xi_{\pm} = \xi_{tt} \pm \xi_{\times\times}$, which can be related to the power spectrum of convergence field by
\begin{equation}
  \xi_{\pm}(\vartheta) = \int^{\infty}_{0}\frac{{\rm d}\ell \ell}{2\pi} J_{0,4}(\ell\vartheta) C^{\kappa\kappa}(\ell),
\end{equation}
where $J_{0,4}(\ell\vartheta)$ denotes the zeroth and fourth Bessel function for $\xi_{+}$ and $\xi_{-}$, respectively. In this section, we validate the weak lensing 2-point statistics by generating a mock catalogue modeled on the photometric redshifts of DES Year-3 measurements (DES-Y3, \citealt{Abbott+2022PhRvD_DESY3}), which provides the largest sky coverage among the currently available dataset of Stage-3 weak lensing surveys. We adopt the same number density as DES-Y3 to populate the mock galaxies and Fig.~\ref{fig_lens3} shows the redshift distributions of the mock catalogue. The test catalogue covers an area of $\sim 4200 \deg^2$, with a number density 5.9 ${\rm arcmin}^2$. Here we do not aim to exactly reproduce the DES-Y3 catalogue. Instead, we construct a mock catalogue with comparable galaxy number density and redshift binning, and directly compare the resulting galaxy shape correlation functions with those from DES-Y3. Fig.~\ref{fig_lens4} shows the tomographic shear correlations $\xi_{\pm}$ from the mock catalogue. We calculate $\xi^{(ij)}_{\pm}$ by the public code {\tt TreeCorr}\footnote{\url{https://github.com/rmjarvis/TreeCorr}} \citep{Jarvis+2015ascl_treecorr}, where the superscript ($ij$) denotes the different redshift bins used for the calculation of the correlation function. 

Firstly, we compare the shear-only correlation (GG, red dots) with the theoretical predictions (solid lines) derived from the halofit model using {\tt Nicaea}\footnote{\url{https://github.com/CosmoStat/nicaea}} \citep{Kilbinger+2009A&A_nicaea}, using the redshift distribution of galaxies in the lightcone. It can be seen that the correlations of lensing shear from the simulation are consistent with the predictions. Due to the presence of intrinsic ellipticity of galaxies in our catalogue, we then measure the ellipticity correlation (GG+II+GI, black circles), which can be directly compared with the observational results from DES-Y3 (blue circles). Overall, the tomographic correlations in the mock catalogue are slightly higher than those in DES-Y3. To quantify the difference between the model and the data, we calculate the reduced $\chi^2$ following the definition in \cite{Wei+2018ApJ} and find the reduced $\chi^2 \sim 1.63$ between our mock data and DES-Y3 measurements \citep{Abbott+2022PhRvD_DESY3}. This difference is consistent with the fact that the underlying cosmology in JT1G simulation \citep{Jiutian_2025arXiv250321368H} is based on Planck results, with $S_8 \equiv \sigma_8(\Omega_m/0.3)^{0.5} = 0.827_{-0.017}^{+0.019}$, which is about $1.5\sigma$ higher than the DES-Y3 constraint of $S_8 = 0.776 \pm 0.017$. In this comparison, we do not account for the effects of redshift uncertainties. These systematics, including shape noise and photometric redshift uncertainties, are incorporated in our CSST data challenge project \footnote{\url{https://nadc.china-vo.org/events/CSSTdatachallenge}} , where more than five types of systematics have been modeled. A more detailed analysis that fully accounts for these effects will be presented in future work.

%, indicating overall good agreement between the mock catalogue and observational catalogue, and meeting the current requirements for the development of data processing.

\section{Summary}
\label{sect:summary}
As one of the Stage-IV galaxy surveys, CSST will use galaxy clustering and weak gravitational lensing to constrain cosmological parameters and unravel the cause of the accelerated expansion of the universe. In current definition stage, it is crucial to construct a mock catalogue, which includes a wide range of realistic features, for optimizing the data processing pipeline and assessing scientific performance. In this paper a mock catalogue has been presented, consisting of stars, galaxies and quasars, for the CSST main survey simulator and could be used to support the analyses of CSST galaxy clustering and weak gravitational lensing. 

We produce star catalogue using either {\tt Galaxia} or {\tt TRILEGAL}. The catalogue provides various properties of each star, including the position, velocity, stellar age, metallicity, chemical information. When using {\tt TRILEGAL}, the catalogue additionally include magnitudes in the CSST photometric system. It is shown that the luminosity function and the mass distribution are generally consistent with observations. 

The mock galaxy catalogue is generated from the halo catalogue using a galaxy formation model of SAM. We utilize JT1G, a cosmological N-body simulation, to construct the merger tree of dark matter haloes, assuming galaxies form at the centers of the dark matter haloes according to analytical prescriptions of relevant physical processes, such as gas cooling, star formation, supernova, and black hole feedback. To characterize galaxy morphology, we use B/T to classify different types of galaxies and assume that the shape of an elliptical galaxy roughly follows that of its host dark matter halo, while the rotation axis of a spiral galaxy is aligned with the spin of its halo. For each galaxy in the mock catalogue, its full-SED is generated using the {\tt STARDUSTER} deep-learning model, trained on {\tt SKIRT} radiative transfer simulations. The magnitudes in each band are then determined with the throughputs of CSST. Additionally, quasars are assigned to galaxies by matching their luminosities with the AGN accretion rates in the galaxy catalogue. The quasar catalogue is constructed using the luminosity functions of \citep{Ross+2013ApJ} and SEDs for quasars are generated with {\tt SIMQSO}.

With the mock galaxy catalogue, we construct a light-cone up to redshift $z \sim 3.5$ over the full sky and perform the weak lensing ray-tracing simulation on the curved sky with a {\tt HEALPix} resolution of $0.43$ arcmin. This enables us to derive the convergence, shear and magnification at each galaxy's position within the light-cone. We validate the mock galaxy catalogue by comparing it with observational data. Our results generally reproduce the stellar mass-halo mass relation at low redshifts but show lower stellar masses for larger halo masses at $z = 2$, consistent with the stellar mass function at that redshift. The size distributions of the mock galaxies show reasonable agreement with observations and our mock catalogue includes more bright galaxies than SDSS data, likely due to the stellar population synthesis used in our data production pipeline. Additionally, we check the weak lensing properties of the mock catalogue. The measured power spectra of convergence and shear E-/B-mode align well with expectations from the Halofit model. As an example of data production pipline, we generate a DES-Y3-like mock catalogue and compare the shear correlation with observational data and halofit predictions, finding good agreement with the data.

Although our mock catalogue includes comprehensive properties, some improvements can be further explored. In the catalogue, galaxies are modeled as a two-component system, bulge and disk. While this is sufficient for some analyses, incorporating more complex morphologies would enhance the catalogue’s applicability, especially for validating data processing pipelines with imaging data. Another potential refinement is the calculation of galaxy SEDs. Currently, SEDs are computed with random incidence angles, which lacks accuracy for individual galaxies. A more precise method would compute SEDs based on the actual position and orientation of each galaxy within the light-cone. Although this is computationally expensive, it could significantly improve the precision of SEDs.

\normalem
\begin{acknowledgements}
This work is supported by the project of the CSST scientific data processing and analysis system of the China Manned Space Project. We thank the working group of CSST-`JiuTian' to make the cosmological simulations available. This work is supported by the NSFC (No. 11903082) and the China Manned Space Program with grant No. CMS-CSST-2025-A20, CMS-CSST-2025-A05, CMSCSST-2021-A03. Y. Chen acknowledges the support by the Natural Science Research Project of Anhui Educational Committee No. 2024AH050049, National Natural Science Foundation of China (NSFC) No. 12003001 and the Anhui Project (Z010118169).

\end{acknowledgements}

\appendix                  %%appendicial material is supported
\section{Statistics of Stellar mass and halo mass}\label{appendixA}
In SAM models, stellar mass is a crucial parameter that can be directly compared with observational data. We present statistical results of the conditional stellar mass function, the evolution of the stellar mass function, and the stellar mass–halo mass relation, comparing them with observational data. Our mock catalogue can reproduce the fundamental relation between stellar mass and halo mass.

\begin{figure*}
    \centering 
    \includegraphics[width=0.8\textwidth]{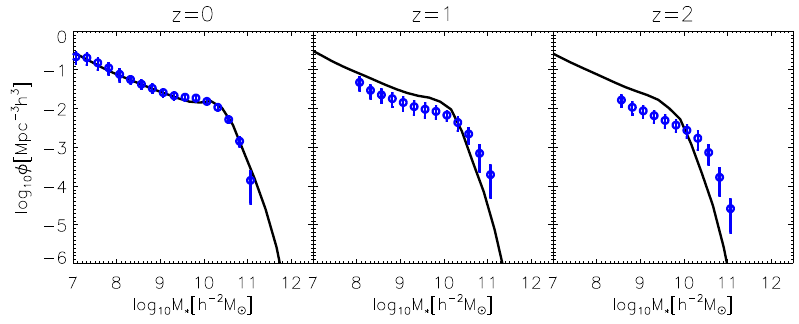} \\
    \includegraphics[width=0.8\textwidth]{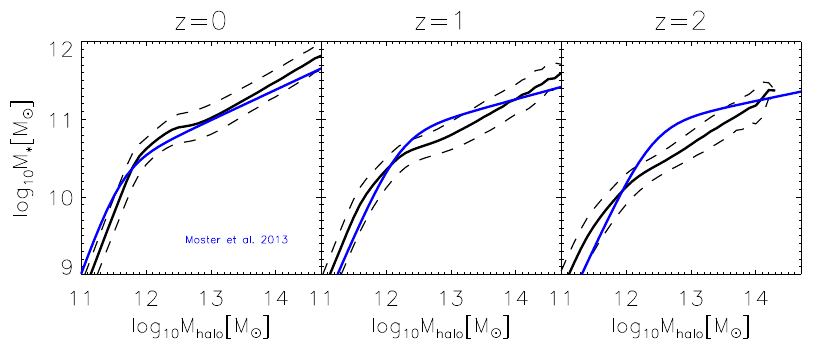}
    \caption{ 
    Upper panels show the evolution of stellar mass functions from $z=2$ to 0. The blue symbols are the combined observation data from \cite{Henriques+2015MNRAS}.
    Lower panels show the evolution of stellar mass-halo mass relations from $z=2$~to 0. Black curves with dashed curves are the results of our galaxy catalogue with 1$\sigma$, blue curves are from the \cite{Moster+2013MNRAS}.
    }
    \label{fig:smf}
\end{figure*}

Fig.~\ref{fig:smf} (upper panels) shows the evolution of stellar mass function of our galaxy catalogue from $z=2$ to 0 (black solid lines). The blue symbols are the combined observational data from \cite{Henriques+2015MNRAS}. The stellar mass of our galaxy catalogue at $z=2$ is slightly lower at high mass end, while higher at low mass end.  

In Fig.~\ref{fig:smf} (lower panels), we show the evolution of stellar mass-halo mass relations from $z=2$ to 0 and compare to the analytical result from \cite{Moster+2013MNRAS}. The stellar mass–halo mass relation is well reproduced by our results at low redshifts. However, at $z=2$, the model predicts lower stellar masses in massive haloes, in agreement with the corresponding stellar mass function.

\begin{figure}
    \centering 
    \includegraphics[width=0.49\textwidth]{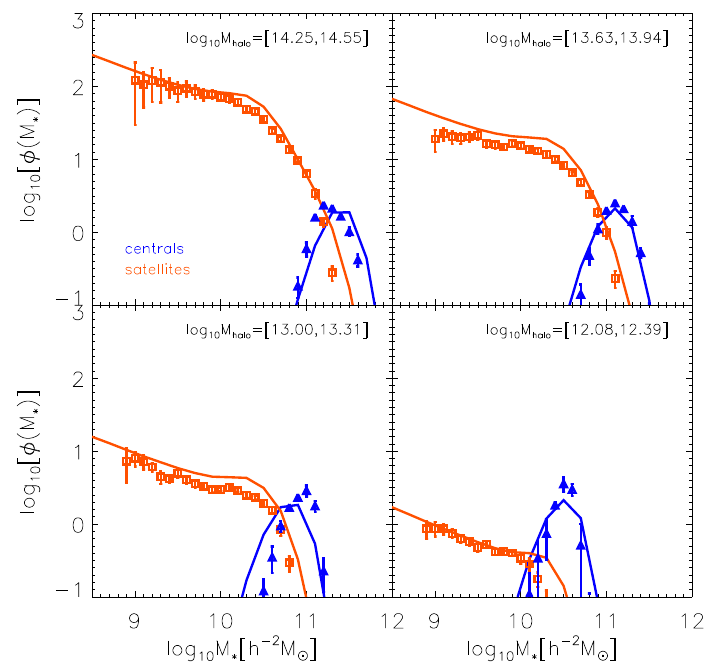}
    \caption{The conditional stellar mass functions at $z=0$, in different halo mass bins. The data points are from \cite{Yang+2012ApJ}. The red/blue lines are for satellites/central galaxies.}
    \label{fig:csmf}
\end{figure}

The conditional stellar mass function indicates the mass distribution of central galaxies and satellite galaxies in certain mass haloes. In Fig. \ref{fig:csmf} we show the conditional stellar mass functions of our galaxy catalogue at $z=0$ which reproduced the observation data \citep{Yang+2012ApJ} especially in massive halo mass bins.


\bibliographystyle{raa}
\bibliography{bibtex}
 
\end{document}